\documentclass[preprint,aps]{revtex4}
\usepackage{axodraw}
\usepackage{amssymb}

\newcommand{\be}{\begin{equation}}
\newcommand{\ee}{\end{equation}}

\newcommand{\reseteqnum}{\setcounter{equation}{0}}
\newcommand{\nn}{\nonumber}

\newcommand{\ovl}[1]{\overline{#1}}
\newcommand{\wt}[1]{\widetilde{#1}}

\newcommand{\eqn}[1]{(\ref{#1})}
\newcommand{\p}{\partial}
\newcommand{\bpsi}{{\overline{\psi}}}

\newcommand{\bq}{{\overline{q}}}
\newcommand{\bu}{{\overline{u}}}
\newcommand{\bd}{{\overline{d}}}
\newcommand{\bs}{{\overline{s}}}

\newcommand{\vev}[1]{\left\langle #1 \right\rangle}

\newcommand{\pslash}{p\kern-1ex /}
\newcommand{\Dslash}{{\cal D}\kern-1.5ex /}
\newcommand{\leftD}{{\mathop{D}\limits^{\leftarrow}}}
\newcommand{\rightD}{{\mathop{D}\limits^{\rightarrow}}}
\newcommand{\tr}{{\rm tr}}

\begin{document}
\begin{flushright}  
{\normalsize UTHEP-638}\\
{\normalsize UTCCS-P-65}\\
\end{flushright}
\title{
Perturbative renormalization factors of four-quark operators for
improved Wilson fermion action and Iwasaki gauge action}

\author{Y.~Taniguchi}
\affiliation{
Graduate School of Pure and Applied Sciences,
University of Tsukuba,
Tsukuba, Ibaraki 305-8571, Japan\\
Center for Computational Physics,
University of Tsukuba,
Tsukuba, Ibaraki 305-8577, Japan
}

%email-address{
%tanigchi@het.ph.tsukuba.ac.jp
%}

\date{\today}

\begin{abstract}
 The renormalization factor and $O(a)$ improvement coefficient of
 four-quark operators are calculated perturbatively for the improved
 Wilson fermion action with clover term and the Iwasaki gauge action.
 With an application to the $K\to\pi\pi$ decay amplitude in mind, the
 calculation is restricted to the parity odd operator, for which the
 operators are multiplicatively renormalized without any mixing to
 operators that have different chiral structures.

\end{abstract}

\maketitle

%\pacs{11.15Ha, 12.38Bx, 12.38Gc}
%11.15Ha : Lattice gauge theory (see also 12.38.G Lattice QCD calculations)
%12.38Bx : Perturbative calculations
%12.38Gc : Lattice QCD calculations (see also 11.15.H Lattice gauge theory)

\reseteqnum
% ------------------------ section 1 ------------------------
\section{Introduction}

Calculation of hadron matrix elements of phenomenological interest
is one of major applications of lattice QCD.
When one tries to calculate weak matrix elements of four quark operators using the
Wilson fermion action, however, one encounters an obstacle since
unwanted mixings with operators having wrong chirality, which are
prohibited in the continuum, is generally introduced through quantum
corrections.

This problem is absent for parity odd operators.
Using the discrete symmetries of parity, charge conjugation and
flavor exchanging transformations,  it was shown \cite{Donini:1999sf}
that the parity odd four quark operator has no extra mixing with wrong
chirality operators even without chiral symmetry.
This is a welcome feature for calculation of the
$K\to\pi\pi$ decay amplitude with the Wilson fermion action.

An improvement with the clover term is indispensable for the Wilson
fermion action.
Since the renormalization group improved gauge action of Iwasaki type 
\cite{Iwasaki:2011jk} has good scaling property already at lattice
spacings around $a^{-1}\sim2$ GeV, the combination of the Iwasaki gauge
action and the improved Wilson fermion action with clover term is a
plausible choice for numerical simulations.
Unfortunately renormalization factors of the four quark operators are
not available for this combination of the actions.
In this paper we calculate the renormalization factor of four
quark operators which contribute to the $K\to\pi\pi$
decay to one-loop order in perturbation theory.

This paper is organized as follows.
In Sec.~\ref{sec:action} we briefly introduce the action and the Feynman
rules relevant for the present calculation.
In Sec.~\ref{sec:4fermi} the $\Delta S=1$ four quark operators are
introduced, which contribute to $K\to\pi\pi$ decay.
The one loop contributions are briefly reviewed in Sec.~\ref{sec:gluon}
for gluon exchange diagrams and are calculated in Sec.~\ref{sec:penguin}
for penguin diagrams.
The renormalization factors are evaluated in
Sec.~\ref{sec:renormalization} for the $\ovl{\rm MS}$ scheme.
Sec.~\ref{sec:oa} is devoted for an evaluation of $O(a)$ effect.
Our conclusion is in Sec.~\ref{sec:conclusion}.

The physical quantities are expressed in lattice units 
and the lattice spacing $a$ is suppressed unless necessary. 
We take SU($N$) gauge group with the gauge coupling $g$ and the second
Casimir $C_F = \displaystyle \frac{N^2-1}{2N}$, while $N=3$ is specified
in the numerical calculations.

\reseteqnum
% ------------------------ section 2 ------------------------
\section{Action and Feynman rules}
\label{sec:action}

We adopt the Iwasaki gauge action
\begin{equation}
S_{\rm gluon} = \frac{1}{g^2}\left\{
c_0 \sum_{plaquette} {\rm Tr}U_{pl}
+ c_1  \sum_{rectangle} {\rm Tr} U_{rtg}
\right\}
\end{equation}
with $c_1=-0.331$ \cite{Iwasaki:2011jk} and the improved Wilson fermion
action with the clover term
\begin{eqnarray}
&&
S_{\rm fermion}=
\sum_n \bpsi(n) \left( \gamma_\mu D_\mu 
- \frac{r}{2} D^2 + m_0 \right) \psi(n)
- c_{\rm SW} \sum_n \sum_{\mu, \nu}
ig \frac{r}{4} \bpsi_n \sigma_{\mu \nu} P_{\mu \nu} (n) \psi_n,
\end{eqnarray}
where $\sigma_{\mu\nu}=\frac{1}{2}\left[\gamma_\mu,\gamma_\nu\right]$.
We shall set $r=1$ in the following.

Weak coupling perturbation theory is developed by expanding the action
in terms of gauge coupling.
The gluon propagator of the Iwasaki action is given as an inverse of the
action kernel
\begin{eqnarray}
&&
G_{\mu\nu}^{AB}(p) =
\left(D^{-1}(p)\right)_{\mu\nu}\delta^{AB},
\\&&
S_{\rm gluon}^{\rm free}=
\frac{1}{2} \int_{-\pi}^{\pi}\frac{d^4p}{(2\pi)^4}
\sum_{\mu, \nu}A_\mu^B(p)D_{\mu\nu}(p)A_\nu^B(-p) , 
\\&&
D_{\mu \nu}(p) = \hat{p}_\mu \hat{p}_\nu
+\sum_\rho (\hat{p}_\rho \delta_{\mu \nu}-\hat{p}_\mu \delta_{\rho\nu})
q_{\mu \rho} \hat{p}_\rho,
\\&&
q_{\mu \nu} = (1-\delta_{\mu\nu})\left(1 -
c_1(\hat{p}_\mu^2 + \hat{p}_\nu^2) \right),
\\&&
\hat{p}_\mu = 2 \sin \frac{p_\mu}{2},
\end{eqnarray}
where we adopted the Feynman gauge.

The quark propagator is given by that of the ordinary Wilson fermion action
\begin{eqnarray}
&&S_{\rm F}(p)=
\frac{ -i \sum_\mu \gamma_\mu \ovl{p}_\mu + W(p)}{\ovl{p}^2 + W(p)^2},
\\&&
\ovl{p}_\mu = \sin p_\mu,
\\&&
W(p)=M+r\sum_{\mu}\left(1-\cos p_\mu\right).
\end{eqnarray}

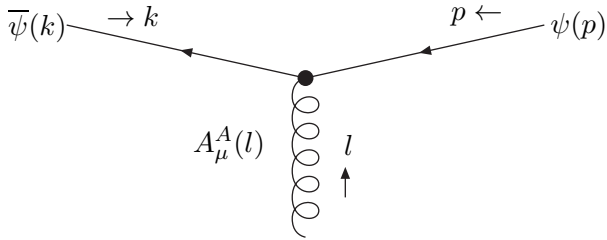
\begin{figure}
\begin{center}\begin{picture}(200,100)(0,0)
\ArrowLine(100,70)(10,90)
\Text(10,90)[r]{$\bpsi(k)$}
\Text(25,90)[lb]{$\rightarrow k$}
\ArrowLine(190,90)(100,70)
\Text(193,90)[l]{$\psi(p)$}
\Text(175,90)[rb]{$p \leftarrow$}
\Gluon(100,10)(100,70){5}{5}
\Text(115,45)[l]{$l$}\LongArrow(115,25)(115,35)
\Text(85,45)[r]{$A^A_\mu(l)$}
\Vertex(100,70){3}
\end{picture}\end{center}
\caption{One gluon interaction vertex.
$k$ and $p$ represent incoming momentum into the vertex.}
\label{fig:vertex}
\end{figure}

We only need single gluon vertexes for our present calculation,
one from the standard gauge coupling
\begin{eqnarray}
V_{1\mu}^A(k,p)
= -i g T^A \left(\gamma_\mu \cos \frac{1}{2}(-k_\mu + p_\mu)
-i r \sin \frac{1}{2}(-k_\mu + p_\mu) \right)
\end{eqnarray}
and another from the clover term
\begin{eqnarray}
V^{(c)A}_{1\mu} (k,p)=
-gT^A c_{\rm SW} \frac{r}{2} \left( \sum_\nu \sigma_{\mu \nu}
\sin (p_\nu + k_\nu) \right) \cos \frac{1}{2} (p_\mu + k_\mu),
\end{eqnarray}
where $k$ and $p$ represent incoming momentum into the vertex as is
shown in Fig.~\ref{fig:vertex}.
$T^A$ $(A=1,\dots,N^2-1)$ is a generator of color SU($N$).

\reseteqnum
% ------------------------ section 3 ------------------------
\section{Four-quark operators}
\label{sec:4fermi}

We shall evaluate the renormalization factor of the following ten
operators
\begin{eqnarray}
&&
Q^{(1)}=\left(\bs d\right)_L\left(\bu u\right)_L,\quad
Q^{(2)}=\left(\bs\times d\right)_L\left(\bu\times u\right)_L,
\\&&
Q^{(3)}=\left(\bs d\right)_L\left(\bu u+\bd d+\bs s\right)_L,\quad
Q^{(4)}=\left(\bs\times d\right)_L\left(\bu\times u+\bd\times d
     +\bs\times s\right)_L,
\\&&
Q^{(5)}=\left(\bs d\right)_L\left(\bu u+\bd d+\bs s\right)_R,\quad
Q^{(6)}=\left(\bs\times d\right)_L\left(\bu\times u+\bd\times d
     +\bs\times s\right)_R,
\\&&
Q^{(7)}=\left(\bs d\right)_L\left(\bu u-\frac{1}{2}\bd d
     -\frac{1}{2}\bs s\right)_R,\quad
Q^{(8)}=\left(\bs\times d\right)_L\left(\bu\times u-\frac{1}{2}\bd\times d
     -\frac{1}{2}\bs\times s\right)_R,
\nn\\
\\&&
Q^{(9)}=\left(\bs d\right)_L\left(\bu u-\frac{1}{2}\bd d
     -\frac{1}{2}\bs s\right)_L,\quad
Q^{(10)}=\left(\bs\times d\right)_L\left(\bu\times u-\frac{1}{2}\bd\times d
     -\frac{1}{2}\bs\times s\right)_L,
\nn\\
\end{eqnarray}
where
\begin{eqnarray}
\left(\bs d\right)_{R/L}=\bs\gamma_\mu\left(1\pm\gamma_5\right)d
\end{eqnarray}
and $\times$ means the following contraction of the color indices
\begin{eqnarray}
Q^{(2)}=\left(\bs\times d\right)_L\left(\bu\times u\right)_L
=\left(\bs_a d_b\right)_L\left(\bu_b u_a\right)_L.
\end{eqnarray}
We notice that these operators are not all independent, satisfying the  relations
\begin{eqnarray}
&&
Q^{(4)}=-Q^{(1)}+Q^{(2)}+Q^{(3)},
\\&&
Q^{(9)}=\frac{1}{2}\left(3Q^{(1)}-Q^{(3)}\right),
\\&&
Q^{(10)}=\frac{1}{2}\left(3Q^{(2)}-Q^{(4)}\right).
\end{eqnarray}

We are interested in the parity odd operators, which contribute to the
$K\to\pi\pi$ decay amplitude
\begin{eqnarray}
&&
Q^{(2n-1)}_{VA+AV}=-Q^{(2n-1)}_{VA}-Q^{(2n-1)}_{AV},
\quad
Q^{(2n)}_{VA+AV}=-Q^{(2n)}_{VA}-Q^{(2n)}_{AV},
\quad (n=1, 2, 5),
\\&&
Q^{(2n-1)}_{VA-AV}=Q^{(2n-1)}_{VA}-Q^{(2n-1)}_{AV},
\quad
Q^{(2n)}_{VA-AV}=Q^{(2n)}_{VA}-Q^{(2n)}_{AV},
\quad (n=3, 4),
\\&&
Q^{(2n-1)}_{VA}=
\left(\bs d\right)_V\sum_{q=u,d,s}\alpha^{(n)}_q\left(\bq q\right)_A,
\\&&
Q^{(2n-1)}_{AV}=
\left(\bs d\right)_A\sum_{q=u,d,s}\alpha^{(n)}_q\left(\bq q\right)_V,
\\&&
Q^{(2n)}_{VA}=
\left(\bs\times d\right)_V
\sum_{q=u,d,s}\alpha^{(n)}_q\left(\bq\times q\right)_A,
\\&&
Q^{(2n)}_{AV}=
\left(\bs\times d\right)_A
\sum_{q=u,d,s}\alpha^{(n)}_q\left(\bq\times q\right)_V,
\end{eqnarray}
where the coefficients are given for $q=u,d,s$
\begin{eqnarray}
&&
\alpha^{(1)}_q=\left(1,0,0\right),
\\&&
\alpha^{(2)}_q=\alpha^{(3)}_q=\left(1,1,1\right),
\\&&
\alpha^{(4)}_q=\alpha^{(5)}_q=\left(1,-\frac{1}{2},-\frac{1}{2}\right)
\label{eqn:alpha-factor}
\end{eqnarray}
and current-current vertex means
\begin{eqnarray}
\left(\bs d\right)_V\left(\bq q\right)_A
=\left(\bs\gamma_\mu d\right)\left(\bq\gamma_\mu\gamma_5q\right).
\end{eqnarray}

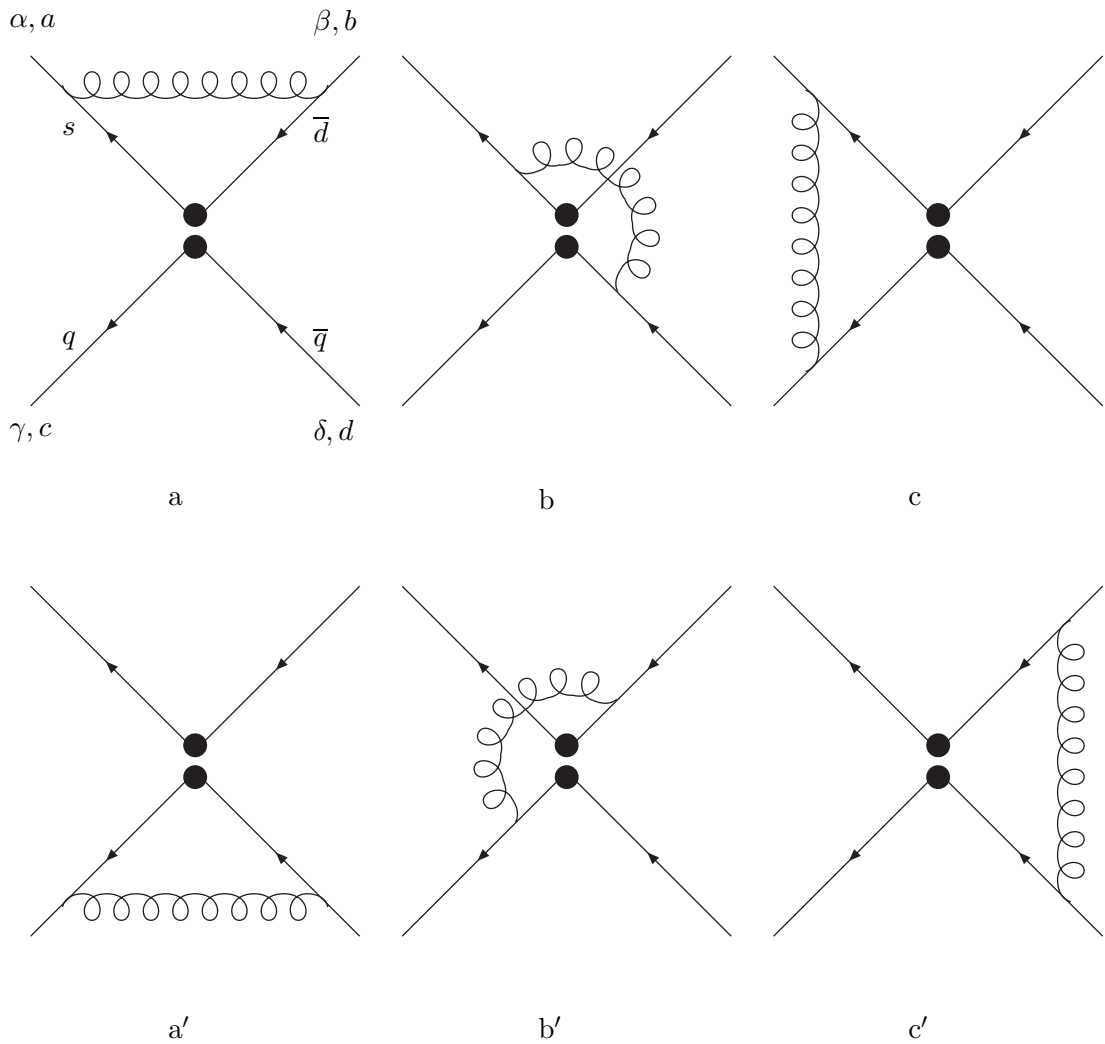
\begin{figure}
\begin{center}
\begin{picture}(420,450)(0,20)
%\put(30,130){Fig.1'}
\Text(60,60)[l]{${\rm a}^\prime$}
\Vertex(70,166){4.5}
\Vertex(70,154){4.5}
\ArrowLine(68,166)(8,226)
\ArrowLine(132,226)(72,166)
\ArrowLine(68,154)(8,94)
\ArrowLine(132,94)(72,154)
\Gluon(20,105)(120,105){5}{8}
%
%\put(30,230){Fig.1}
\Text(60,260)[l]{a}
\Text(0,440)[l]{$\alpha,a$}
\Text(115,440)[l]{$\beta,b$}
\Text(0,285)[l]{$\gamma,c$}
\Text(115,285)[l]{$\delta,d$}
\Text(20,400)[l]{$s$}
\Text(115,400)[l]{$\ovl{d}$}
\Text(20,320)[l]{$q$}
\Text(115,320)[l]{$\bq$}
\Vertex(70,366){4.5}
\Vertex(70,354){4.5}
\ArrowLine(68,366)(8,426)
\ArrowLine(132,426)(72,366)
\ArrowLine(68,354)(8,294)
\ArrowLine(132,294)(72,354)
\Gluon(20,415)(120,415){-5}{8}
%
%\put(170,130){Fig.2}
\Text(200,60)[l]{${\rm b}^\prime$}
\Vertex(210,166){4.5}
\Vertex(210,154){4.5}
\ArrowLine(208,166)(148,226)
\ArrowLine(272,226)(212,166)
\ArrowLine(208,154)(148,94)
\ArrowLine(272,94)(212,154)
\GlueArc(210,160)(30,51,230){-5}{7}
%
%\put(170,230){Fig.2}
\Text(200,260)[l]{b}
\Vertex(210,366){4.5}
\Vertex(210,354){4.5}
\ArrowLine(208,366)(148,426)
\ArrowLine(272,426)(212,366)
\ArrowLine(208,354)(148,294)
\ArrowLine(272,294)(212,354)
\GlueArc(210,360)(30,-50,130){-5}{7}
%
%\put(100,130){Fig.3}
\Text(340,60)[l]{${\rm c}^\prime$}
\Vertex(350,166){4.5}
\Vertex(350,154){4.5}
\ArrowLine(348,166)(288,226)
\ArrowLine(412,226)(352,166)
\ArrowLine(348,154)(288,94)
\ArrowLine(412,94)(352,154)
\Gluon(400,213)(400,107){-5}{8}
%
%\put(100,230){Fig.3}
\Text(340,260)[l]{c}
\Vertex(350,366){4.5}
\Vertex(350,354){4.5}
\ArrowLine(348,366)(288,426)
\ArrowLine(412,426)(352,366)
\ArrowLine(348,354)(288,294)
\ArrowLine(412,294)(352,354)
\Gluon(300,413)(300,307){5}{8}
\end{picture}
\end{center}
\caption{One-loop vertex corrections for the four-quark operator
(gluon exchanging diagram).
$\alpha,\beta,\gamma,\delta$ and $a,b,c,d$ label Dirac and color
indices respectively.}
\label{fig:diagrams}
\end{figure}

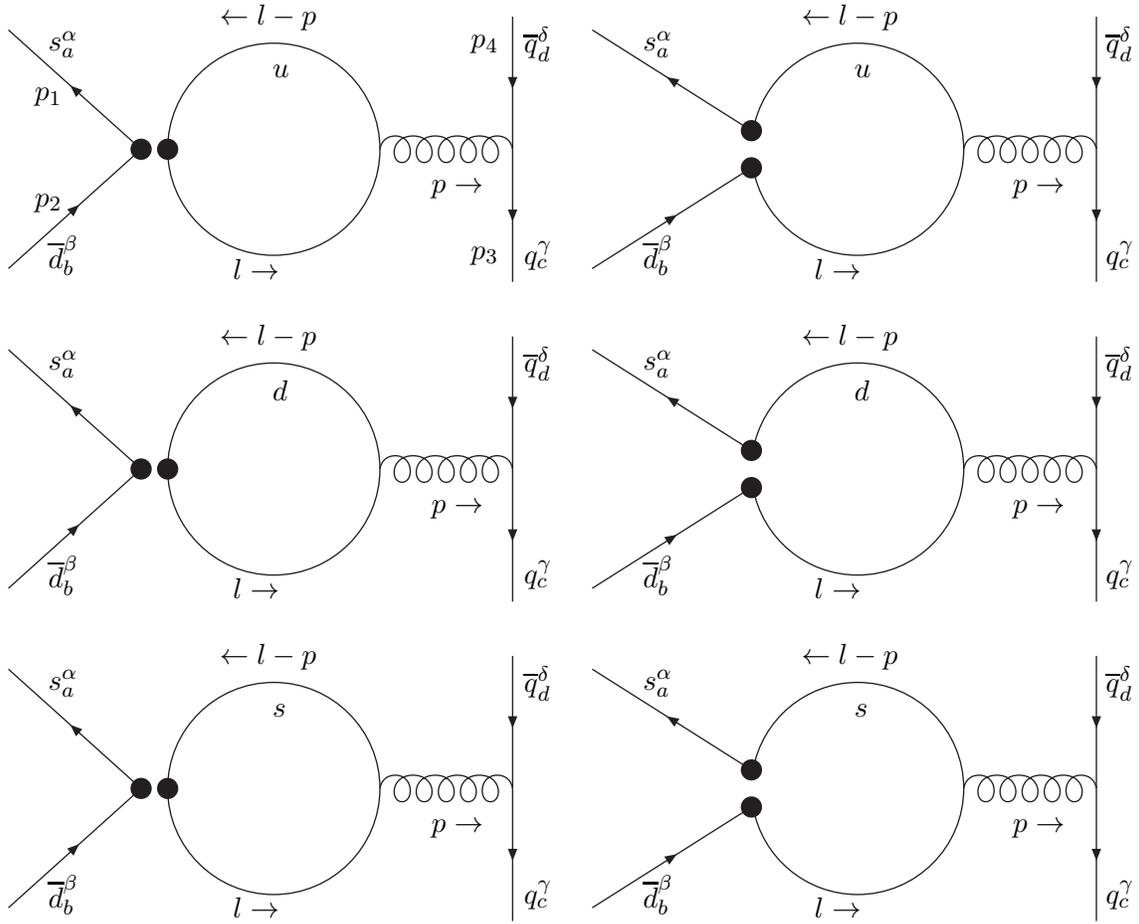
\begin{figure}
\begin{center}
\begin{picture}(420,120)(0,0)
%\Text(50,120)[l]{Penguin diagram}
\Text(15,90)[l]{$s_a^\alpha$}
\Text(10,70)[l]{$p_1$}
\Text(10,30)[l]{$p_2$}
\Text(15,10)[l]{$\ovl{d}_b^\beta$}
\Text(195,90)[l]{$\ovl{q}_d^\delta$}
\Text(175,90)[l]{$p_4$}
\Text(175,10)[l]{$p_3$}
\Text(195,10)[l]{$q_c^\gamma$}
\Text(100,80)[l]{$u$}
\Text(85,5)[l]{$l \rightarrow$}
\Text(80,100)[l]{$\leftarrow l-p$}
\Text(160,35)[l]{$p\rightarrow$}
\Vertex(50,50){4}
\Vertex(60,50){4}
\ArrowLine(0,5)(50,50)
\ArrowLine(50,50)(0,95)
\CArc(100,50)(40,0,360)
\Gluon(140,50)(190,50){5}{5}
\ArrowLine(190,50)(190,0)
\ArrowLine(190,100)(190,50)
\Text(240,90)[l]{$s_a^\alpha$}
\Text(240,10)[l]{$\ovl{d}_b^\beta$}
\Text(415,90)[l]{$\ovl{q}_d^\delta$}
\Text(415,10)[l]{$q_c^\gamma$}
\Text(320,80)[l]{$u$}
\Text(305,5)[l]{$l \rightarrow$}
\Text(300,100)[l]{$\leftarrow l-p$}
\Text(380,35)[l]{$p\rightarrow$}
\Vertex(280,43){4}
\Vertex(280,57){4}
\ArrowLine(220,5)(280,43)
\ArrowLine(280,57)(220,95)
\CArc(320,50)(40,190,170)
\Gluon(360,50)(410,50){5}{5}
\ArrowLine(410,50)(410,0)
\ArrowLine(410,100)(410,50)
\end{picture}
%%%%%%%%%%%%%%%
\begin{picture}(420,120)(0,0)
\Text(15,90)[l]{$s_a^\alpha$}
\Text(15,10)[l]{$\ovl{d}_b^\beta$}
\Text(195,90)[l]{$\ovl{q}_d^\delta$}
\Text(195,10)[l]{$q_c^\gamma$}
\Text(100,80)[l]{$d$}
\Text(85,5)[l]{$l \rightarrow$}
\Text(80,100)[l]{$\leftarrow l-p$}
\Text(160,35)[l]{$p\rightarrow$}
\Vertex(50,50){4}
\Vertex(60,50){4}
\ArrowLine(0,5)(50,50)
\ArrowLine(50,50)(0,95)
\CArc(100,50)(40,0,360)
\Gluon(140,50)(190,50){5}{5}
\ArrowLine(190,50)(190,0)
\ArrowLine(190,100)(190,50)
\Text(240,90)[l]{$s_a^\alpha$}
\Text(240,10)[l]{$\ovl{d}_b^\beta$}
\Text(415,90)[l]{$\ovl{q}_d^\delta$}
\Text(415,10)[l]{$q_c^\gamma$}
\Text(320,80)[l]{$d$}
\Text(305,5)[l]{$l \rightarrow$}
\Text(300,100)[l]{$\leftarrow l-p$}
\Text(380,35)[l]{$p\rightarrow$}
\Vertex(280,43){4}
\Vertex(280,57){4}
\ArrowLine(220,5)(280,43)
\ArrowLine(280,57)(220,95)
\CArc(320,50)(40,190,170)
\Gluon(360,50)(410,50){5}{5}
\ArrowLine(410,50)(410,0)
\ArrowLine(410,100)(410,50)
\end{picture}
%%%%%%%%%%%%%%%
\begin{picture}(420,120)(0,0)
\Text(15,90)[l]{$s_a^\alpha$}
\Text(15,10)[l]{$\ovl{d}_b^\beta$}
\Text(195,90)[l]{$\ovl{q}_d^\delta$}
\Text(195,10)[l]{$q_c^\gamma$}
\Text(100,80)[l]{$s$}
\Text(85,5)[l]{$l \rightarrow$}
\Text(80,100)[l]{$\leftarrow l-p$}
\Text(160,35)[l]{$p\rightarrow$}
\Vertex(50,50){4}
\Vertex(60,50){4}
\ArrowLine(0,5)(50,50)
\ArrowLine(50,50)(0,95)
\CArc(100,50)(40,0,360)
\Gluon(140,50)(190,50){5}{5}
\ArrowLine(190,50)(190,0)
\ArrowLine(190,100)(190,50)
\Text(240,90)[l]{$s_a^\alpha$}
\Text(240,10)[l]{$\ovl{d}_b^\beta$}
\Text(415,90)[l]{$\ovl{q}_d^\delta$}
\Text(415,10)[l]{$q_c^\gamma$}
\Text(320,80)[l]{$s$}
\Text(305,5)[l]{$l \rightarrow$}
\Text(300,100)[l]{$\leftarrow l-p$}
\Text(380,35)[l]{$p\rightarrow$}
\Vertex(280,43){4}
\Vertex(280,57){4}
\ArrowLine(220,5)(280,43)
\ArrowLine(280,57)(220,95)
\CArc(320,50)(40,190,170)
\Gluon(360,50)(410,50){5}{5}
\ArrowLine(410,50)(410,0)
\ArrowLine(410,100)(410,50)
\end{picture}
\end{center}
\caption{One-loop vertex corrections for the four-quark operator
(penguin diagram).
$\alpha,\beta,\gamma,\delta$ and $a,b,c,d$ label Dirac and color
indices respectively.
All external momentum $p_i$'s are in-coming direction.}
\label{fig:penguin}
\end{figure}

There are two kinds of one loop corrections to these
operators.
One is given by gluon exchanging diagrams in Fig.~\ref{fig:diagrams} and
the other by the penguin diagrams in Fig.~\ref{fig:penguin}.
The gluon exchanging diagrams preserve the flavor structure and there
occur no mixing between operators with different $n$.
On the other hand the penguin diagrams mix any operator to the penguin
operator unless it belongs to a different representation of the flavor
$SU(3)_f$.

There are a number of parity odd operators having different
chirality
\begin{eqnarray}
&&
O^{(2n-1)}_{SP\pm PS}=
\left(\bs d\right)_S\sum_{q=u,d,s}\alpha^{(n)}_q\left(\bq q\right)_P
\pm\left(\bs d\right)_P\sum_{q=u,d,s}\alpha^{(n)}_q\left(\bq q\right)_S,
\label{eqn:SPPS}
\\&&
O^{(2n)}_{SP\pm PS}=
\left(\bs\times d\right)_S\sum_{q=u,d,s}\alpha^{(n)}_q\left(\bq\times q\right)_P
\pm\left(\bs\times d\right)_P
\sum_{q=u,d,s}\alpha^{(n)}_q\left(\bq\times q\right)_S,
\\&&
O^{(2n-1)}_{T\wt{T}}=
\left(\bs d\right)_T\sum_{q=u,d,s}\alpha^{(n)}_q\left(\bq q\right)_{\wt{T}},
\\&&
O^{(2n)}_{T\wt{T}}=\left(\bs\times d\right)_T
\sum_{q=u,d,s}\alpha^{(n)}_q\left(\bq\times q\right)_{\wt{T}},
\\&&
\left(\bs d\right)_S\left(\bq q\right)_P=
\left(\bs d\right)\left(\bq\gamma_5q\right),
\label{eqn:TTtilde}
\quad
\left(\bs d\right)_T\left(\bq q\right)_{\wt{T}}
=\left(\bs\sigma_{\mu\nu}d\right)\left(\bq\sigma_{\mu\nu}\gamma_5q\right)
\end{eqnarray}
Chiral symmetry does not prohibit mixings of these operators in the
Wilson fermion system.
However, it was shown in Ref.~\cite{Donini:1999sf} that the Wilson fermion system has sufficient set of discrete symmetries to protect the parity odd
operators $Q_{VA\pm AV}$ from such mixings. 

For the gluon exchanging diagrams, the operator mixing can be studied
with the following operator having four flavors
\begin{eqnarray}
&&
O^{(o)}_{\Gamma\Gamma'\pm\Gamma'\Gamma}=
\left(\bpsi_1\psi_2\right)_\Gamma\left(\bpsi_3\psi_4\right)_{\Gamma'}
\pm\left(\bpsi_1\psi_2\right)_{\Gamma'}\left(\bpsi_3\psi_4\right)_{\Gamma},
\\&&
O^{(e)}_{\Gamma\Gamma'\pm\Gamma'\Gamma}=
\left(\bpsi_1\times\psi_2\right)_\Gamma
\left(\bpsi_3\times\psi_4\right)_{\Gamma'}
\pm\left(\bpsi_1\times\psi_2\right)_{\Gamma'}
\left(\bpsi_3\times\psi_4\right)_{\Gamma}.
\end{eqnarray}
It was proved in Ref.~\cite{Donini:1999sf} that parity, charge
conjugation and two types of flavor exchanging transformations
\begin{eqnarray}
&&
{\cal S}' =\left(\psi_1\leftrightarrow\psi_2,\psi_3\leftrightarrow\psi_4\right)
,\quad
{\cal S}''=\left(\psi_1\leftrightarrow\psi_4,\psi_2\leftrightarrow\psi_3\right)
\end{eqnarray}
are sufficient to show that the mixings occur only between $O^{(o)}_{VA+AV}$ and
$O^{(e)}_{VA+AV}$,  or between $O^{(o)}_{VA-AV}$ and $O^{(e)}_{VA-AV}$.
We notice that the Fierz rearrangement leads to relations between the
Fierz partners
\begin{eqnarray}
&&
O^{(e)}_{VA-AV}=2O^{(o)F}_{SP-PS},\quad
O^{(o)}_{VA-AV}=2O^{(e)F}_{SP-PS},
\\&&
O^{(o)F}_{SP-PS}=
\left(\bpsi_1\psi_4\right)_S\left(\bpsi_3\psi_2\right)_P
-\left(\bpsi_1\psi_4\right)_P\left(\bpsi_3\psi_2\right)_S.
\end{eqnarray}
Thus $O_{VA-AV}$ and $O_{SP-PS}$ are basically the same operator rearranged
with each other,  and if we include the Fierz partner $O_{VA-AV}^F$ it
also mixes with $O_{SP-PS}$.

For the penguin diagram we need to keep the three flavors structure. Hence flavor exchange of
\begin{eqnarray}
&&
{\cal S}' =\left(d\leftrightarrow s,\bd\leftrightarrow\bs\right)
\end{eqnarray}
is a symmetry.  
Together with parity and charge conjugation one can show that
the mixing occurs only among $Q_{VA\pm AV}$'s.
%ここの議論がよくわからない。ここのS'は、III.28のS'と違う？同じ？

We note that these discrete symmetries still allow mixings with lower
dimensional parity odd operators
\begin{eqnarray}
&&
\left(m_d-m_s\right)\left(\bs\gamma_5d\right),
\label{eqn:pseudo-scalar}
\\&&
\left(m_d-m_s\right)\p_\mu\left(\bs\gamma_\mu\gamma_5d\right),
\\&&
\left(m_d-m_s\right)\left(\bs F_{\mu\nu}\sigma_{\mu\nu}\gamma_5d\right),
\\&&
\left(m_d-m_s\right)\left(\bs \wt{F}_{\mu\nu}\sigma_{\mu\nu}d\right)
\end{eqnarray}
proportional to a mass difference $\left(m_d-m_s\right)$.

\reseteqnum
% ------------------------ section 4 ------------------------
\section{One loop correction from gluon exchanging diagrams}
\label{sec:gluon}

We consider the following four fermi operator
\begin{eqnarray}
O^{(k)}_{XY}=
\left(T^{(k)}\right)_{ab;cd}
\left(\Gamma_X\otimes\Gamma_Y\right)_{\alpha\beta;\gamma\delta}
\left(\bs_{a,\alpha}d_{b,\beta}\right)
\left(\bq_{c,\gamma}q'_{d,\delta}\right),
\end{eqnarray}
in order to evaluate the one loop correction from the gluon exchanging
diagrams given in Fig.~\ref{fig:diagrams}, where $T$ represents the
color factor
\begin{eqnarray}
&&
\left(T^{(2n-1)}\right)_{ab;cd}
=\left(1\wt{\otimes}1\right)_{ab;cd}
=\delta_{ab}\delta_{cd}\quad
({\rm for}\; Q^{(2n-1)}),
\\&&
\left(T^{(2n)}\right)_{ab;cd}
=\left(1\wt{\odot}1\right)_{ab;cd}
=\delta_{ad}\delta_{bc}
\quad
({\rm for}\; Q^{(2n)}).
\label{eqn:color-factor}
\end{eqnarray}
and $\Gamma$ is the gamma matrix
\begin{eqnarray}
&&
\left(\Gamma_X\otimes\Gamma_Y\right)_{\alpha\beta;\gamma\delta}
=\left(\Gamma_X\right)_{\alpha\beta}\left(\Gamma_Y\right)_{\gamma\delta},
\\&&
\Gamma_V=\gamma_\mu,\quad
\Gamma_A=\gamma_\mu\gamma_5
\end{eqnarray}
where summation over $\mu$ is taken.

Since the one loop correction has already been evaluated in
Ref.~\cite{Constantinou:2010zs} for various gauge actions, we just
briefly review the result for the Iwasaki gauge action.
We consider one loop corrections to the amputated four quark vertex
\begin{eqnarray}
I_{k;XY}=
\vev{O^{(k)}_{XY} s_{a\alpha}\bd_{b\beta}q_{c\gamma}\bq'_{d\delta}
}_{\rm 1PI}.
\end{eqnarray}
The contributions from the diagrams $(a)$ and $(a')$ are given by
\begin{eqnarray}
I^{(a)}_{k;XY} &=& J^{(a)}_k\int_{-\pi}^\pi\frac{d^4l}{(2\pi)^4}
V_{1\mu}(0,l)S_F(l)\Gamma_XS_F(l)V_{1\nu}(-l,0)
\otimes \Gamma_Y G_{\mu\nu}(l),
\\
I^{(a')}_{k;XY} &=& J^{(a)}_k\Gamma_X \otimes
\int_{-\pi}^\pi\frac{d^4l}{(2\pi)^4}
V_{1\mu}(0,l)S_F(l)\Gamma_YS_F(l)V_{1\nu}(-l,0)G_{\mu\nu}(l).
\end{eqnarray}
The contributions from the diagrams $(b)$ and $(b')$ are
\begin{eqnarray}
I^{(b)}_{k;XY} &=& J^{(b)}_k\int_{-\pi}^\pi\frac{d^4l}{(2\pi)^4}
V_{1\mu}(0,l) S_F(l) \Gamma_X
\otimes \Gamma_Y S_F(l) V_{1\nu}(-l,0) G_{\mu\nu}(l),
\\
I^{(b')}_{k;XY} &=& J^{(b)}_k\int_{-\pi}^\pi\frac{d^4l}{(2\pi)^4}
\Gamma_X S_F(l) V_{1\mu}(-l,0)
\otimes V_{1\nu}(0,l) S_F(l) \Gamma_Y G_{\mu\nu}(l).
\end{eqnarray}
The contributions from the diagrams $(c)$ and $(c')$ are
\begin{eqnarray}
I^{(c)}_{k;XY} &=& J^{(c)}_k\int_{-\pi}^\pi\frac{d^4l}{(2\pi)^4}
V_{1\mu}(0,l)S_F(l)\Gamma_X
\otimes V_{1\nu}(0,-l)S_F(-l)\Gamma_Y G_{\mu\nu}(l),
\\
I^{(c')}_{k;XY} &=& J^{(c)}_k\int_{-\pi}^\pi\frac{d^4l}{(2\pi)^4}
\Gamma_X S_F(l)V_{1\mu}(-l,0)
\otimes \Gamma_YS_F(-l)V_{1\nu}(l,0) G_{\mu\nu}(l).
\end{eqnarray}
The color index contributions are already factored out in the above
\begin{eqnarray}
&&
J_{{2n-1}}^{(a)} = T^A 1 T^B \delta^{AB} \wt{\otimes} 1
=C_F 1\wt{\otimes}1,
\\&&
J_{{2n}}^{(a)} = T^A 1 \wt{\odot} 1 T^B \delta^{AB}
=\frac{1}{2}1\wt{\otimes}1-\frac{1}{2N}1\wt{\odot}1,
\\&&
J_{{2n-1}}^{(b)} = T^A 1\wt{\otimes}1 T^A
=\frac{1}{2}1\wt{\odot}1-\frac{1}{2N}1\wt{\otimes}1,
\\&&
J_{{2n}}^{(b)} = T^A 1 T^A\wt{\odot}1=C_F 1\wt{\odot}1,
\\&&
J_{{2n-1}}^{(c)} = T^A 1\wt{\otimes}T^A 1
=\frac{1}{2}1\wt{\odot}1-\frac{1}{2N}1\wt{\otimes}1,
\\&&
J_{{2n}}^{(c)} = T^A 1\wt{\odot}T^A 1
=\frac{1}{2}1\wt{\otimes}1-\frac{1}{2N}1\wt{\odot}1.
\end{eqnarray}

Note that contributions should also be included where one or more of the
gluon vertexes is replaced with $V^{(c)}_{1\mu}$ from the clover term.
We shall adopt an on-shell massless scheme in this section; 
quark mass and all external momenta are set to zero.

\subsection{Contribution from diagram (a) and (a')}

The one loop corrections $I^{(a)}_{XY}$ and $I^{(a')}_{XY}$ are
the same as those to the bilinear (axial) vector current operator.
Omitting the color factor the one loop contribution is given by
\begin{eqnarray}
&&
I_{VA}^{(a)} = T_V\left(V \otimes A\right),\quad
I_{VA}^{(a')} = T_A\left(V \otimes A\right),
\\&&
I_{AV}^{(a)} = T_A\left(A \otimes V\right),\quad
I_{AV}^{(a')} = T_V\left(A \otimes V\right),
\end{eqnarray}
where $V\otimes A=\gamma_\mu\otimes\gamma_\mu\gamma_5$ and
$T_{V/A}$ is the one loop correction to the local (axial) vector
current for the Wilson fermion
\begin{eqnarray}
T_\Gamma\Gamma=
\int_{-\pi}^{\pi}\frac{d^4l}{(2\pi)^4}
V_{1\mu}(0,l)S_F(l)\Gamma S_F(l)V_{1\nu}(-l,0)G_{\mu\nu}(l).
\label{eqn:current-correction}
\end{eqnarray}

Introducing the gluon mass $\lambda$
to the propagator $G_{\mu\nu}(l)$ in the loop we obtain \cite{Aoki:1998ar}
\begin{eqnarray}
T_\Gamma=\frac{g^2}{16\pi^2}\left(-\frac{h_2(\Gamma)}{4}
\ln\left(\lambda a\right)^2+V_\Gamma
\right)
\label{eqn:lattice-scheme}
\end{eqnarray}
in the Feynman gauge, where $h_2(\Gamma)$ is an integer given by
\begin{equation}
h_2(\Gamma)=4(A),\; 4(V),\; 16(P),\; 16(S),\; 0(T)
\end{equation}
for various Dirac channels.
The finite constants 
$V_\Gamma$ depend quadratically on the clover coefficients $c_{\rm SW}$,
and we write
\begin{eqnarray}
V_\Gamma &=& V_\Gamma^{(0)}+c_{\rm SW} V_\Gamma^{(1)}
+c_{\rm SW}^2 V_\Gamma^{(2)}.
\end{eqnarray}
The superscript $(i=0,1,2)$ means a correction of $i$-th order in
$c_{\rm SW}$, where $i$ gauge interactions are replaced with that from
the clover term.
The numerical value of the finite part $V_\Gamma$ has already been
evaluated in Ref.~\cite{Aoki:1998ar} for various gauge actions and is
given in Table \ref{tab:local} for Iwasaki gauge action.
%We notice that no mixing appears from this diagram.

\subsection{Contribution from diagram (b) and (b')}

The contributions from the diagrams $(b)$ and $(b')$ can be evaluated by
using the Fierz rearrangement \cite{Martinelli:1983ac} in the spinor
indices.

Each one loop correction turned out to be the same as
that to the (pseudo) scalar density and (axial) vector current operators
given by \eqn{eqn:current-correction}.
%\begin{eqnarray}
%I^{(b)}_{VA+AV} &=&
%-\int\left( V_{1\lambda}(0,l) S_F(l) \gamma_\mu S_F(l) V_{1\nu}(-l,0)\right)
%\odot \left(\gamma_\mu\gamma_5\right) G_{\lambda \nu}(l)
%\nn\\&&
%-\int\left( V_{1\lambda}(0,l) S_F(l) \gamma_\mu\gamma_5 S_F(l) V_{1\nu}(-l,0)
%\right)
%\odot \left( \gamma_\mu \right) G_{\lambda \nu}(l),
%\\
%I^{(b)}_{VA-AV} &=&
%-2\int\left(V_{1\lambda}(0,l) S_F(l) S_F(l) V_{1\nu}(-l,0) \right)
%\odot \left(\gamma_5\right) G_{\lambda \nu}(l)
%\nn\\&&
%+2\int\left(V_{1\lambda}(0,l) S_F(l)\gamma_5S_F(l) V_{1\nu}(-l,0) \right)
%\odot \left(1\right) G_{\lambda \nu}(l),
%\\
%I^{(b')}_{VA+AV} &=&
%-\int\left(\gamma_\mu\right)\odot
%\left(V_{1\nu}(0,l) S_F(l) \gamma_\mu\gamma_5 S_F(l) V_{1\lambda}(-l,0)\right)
%G_{\lambda \nu}(l),
%\nn\\&&
%-\int\left(\gamma_\mu\gamma_5\right)\odot
%\left(V_{1\nu}(0,l) S_F(l) \gamma_\mu S_F(l) V_{1\lambda}(-l,0) \right)
%G_{\lambda \nu}(l),
%\\
%I^{(b')}_{VA-AV} &=&
%-2\int\left(1\right)\odot
%\left(V_{1\nu}(0,l) S_F(l) \gamma_5 S_F(l) V_{1\lambda}(-l,0)\right)
%G_{\lambda \nu}(l),
%\nn\\&&
%+2\int\left(\gamma_5\right)\odot
%\left(V_{1\nu}(0,l) S_F(l) S_F(l) V_{1\lambda}(-l,0)\right)
%G_{\lambda \nu}(l),
%\end{eqnarray}
After carrying out the loop integral we obtain
\begin{eqnarray}
I_{VA+AV}^{(b)}+I_{VA+AV}^{(b')} &=&
-\left(T_V+T_A\right)\left(V \odot A+A \odot V\right),
\\
I_{VA-AV}^{(b)}+I_{VA-AV}^{(b')} &=&
-2\left(T_S+T_P\right)\left(S \odot P-P \odot S\right),
\end{eqnarray}
where the direct product $\odot$ means
\begin{eqnarray}
\left(\Gamma\odot\Gamma'\right)_{\alpha\beta;\gamma\delta}
=\left(\Gamma\right)_{\alpha\delta}\left(\Gamma'\right)_{\gamma\beta}
\end{eqnarray}
and $S \odot P=1\odot\gamma_5$.
Performing the Fierz rearrangement again the vertex function is
transformed into the same spinor structure as at the tree level
\begin{eqnarray}
I_{VA+AV}^{(b)}+I_{VA+AV}^{(b')} &=&
\left(T_V+T_A\right)\left(V \otimes A+A \otimes V\right),
\\
I_{VA-AV}^{(b)}+I_{VA-AV}^{(b')} &=&
\left(T_S+T_P\right)\left(V \otimes A-A \otimes V\right).
\end{eqnarray}

\subsection{Contribution from diagram (c) and (c')}

Evaluation of the contributions from the diagrams $(c)$, $(c')$ is performed
by using charge conjugation and Fierz rearrangement
\cite{Martinelli:1983ac}.
We use the representation of the charge conjugation matrix 
${C} = \gamma_2 \gamma_0$
and the relations 
\begin{eqnarray}
&&
%{C}^{-1} \gamma_\mu {C}={C} \gamma_\mu {C}^{-1} = -\gamma_\mu^T,
%\\&&
CS_{\rm F}(p)C^{-1}=S_{\rm F}(-p)^T,
\\&&
CV_{1\mu}(k,p)C^{-1}=-V_{1\mu}(p,k)^T,
\\&&
CV^{(c)}_{1\mu}(k,p)C^{-1}=-V^{(c)}_{1\mu}(p,k)^T.
\end{eqnarray}

Using these relations
%, the one loop correction is re-written as follows
%\begin{eqnarray}
%I^{(c)}_{VA} &=&\int
%\left[
%V_{1\lambda}(0,l) S_F(l) \gamma_\mu
%\otimes
%C\gamma_\mu\gamma_5S_F(l)\left(-V_{1\nu}(-l,0)\right)
%C^{-1}\right]_{\alpha\beta;\delta\gamma}
%G_{\lambda \nu}(l),
%\\
%I^{(c)}_{AV} &=&\int
%\left[V_{1\lambda}(0,l) S_F(l) \gamma_\mu\gamma_5
%\otimes
%C\gamma_\mu S_F(l)V_{1\nu}(-l,0)C^{-1}\right]_{\alpha\beta;\delta\gamma}
%G_{\lambda \nu}(l),
%\\
%I^{(c')}_{VA} &=&\int
%\left[\gamma_\mu S_F(l) V_{1\lambda}(-l,0)
%\otimes
%C\left(-V_{1\nu}(0,l)\right)S_F(l)\gamma_\mu\gamma_5
%C^{-1} \right]_{\alpha\beta;\delta\gamma}
%G_{\lambda \nu}(l),
%\\
%I^{(c')}_{AV} &=&\int
%\left[\gamma_\mu\gamma_5 S_F(l) V_{1\lambda}(-l,0)
%\otimes
%C\left(-V_{1\nu}(0,l)\right)S_F(l)\left(-\gamma_\mu\right)
%C^{-1} \right]_{\alpha\beta;\delta\gamma}
%G_{\lambda \nu}(l).
%\end{eqnarray}
%Then we use
and the Fierz rearrangement
%\begin{eqnarray}
%I^{(c)}_{VA+AV} &=&
%2\int\left[V_{1\lambda}(0,l) S_F(l)S_F(l) V_{1\nu}(-l,0){\cal C}^{-1}
%\odot {\cal C} \gamma_5 \right]_{\alpha\beta;\delta\gamma}
%G_{\lambda \nu}(l)
%\nn\\&-&
%2\int\left[ V_{1\lambda}(0,l) S_F(l)\gamma_5S_F(l) V_{1\nu}(-l,0){\cal C}^{-1}
%\odot {\cal C} \right]_{\alpha\beta;\delta\gamma}
%G_{\lambda \nu}(l),
%\\
%I^{(c)}_{VA-AV} &=&
%\int\left[V_{1\lambda}(0,l) S_F(l) \gamma_\mu S_F(l) V_{1\nu}(-l,0){\cal C}^{-1}
%\odot {\cal C} \gamma_\mu\gamma_5 \right]_{\alpha\beta;\delta\gamma}
%G_{\lambda \nu}(l)
%\nn\\&+&
%\int\left[V_{1\lambda}(0,l) S_F(l)\gamma_\mu\gamma_5 S_F(l) V_{1\nu}(-l,0)
%{\cal C}^{-1}
%\odot {\cal C} \gamma_\mu\right]_{\alpha\beta;\delta\gamma}
%G_{\lambda \nu}(l),
%\\
%I^{(c')}_{VA+AV} &=&
%2\int\left[ C^{-1}\odot
%CV_{1\nu}(0,l)S_F(l)\gamma_5S_F(l) V_{1\lambda}(-l,0)
%\right]_{\alpha\beta;\delta\gamma}
%G_{\lambda \nu}(l)
%\nn\\&-&
%2\int\left[\gamma_5C^{-1}\odot
%CV_{1\nu}(0,l)S_F(l)S_F(l) V_{1\lambda}(-l,0)\right]_{\alpha\beta;\delta\gamma}
%G_{\lambda \nu}(l),
%\\
%I^{(c')}_{VA-AV} &=&
%\int\left[\gamma_\mu C^{-1}\odot
%CV_{1\nu}(0,l)S_F(l)\gamma_\mu\gamma_5S_F(l) V_{1\lambda}(-l,0)
%\right]_{\alpha\beta;\delta\gamma}
%G_{\lambda \nu}(l)
%\nn\\&+&
%\int\left[\gamma_\mu\gamma_5C^{-1}\odot
%CV_{1\nu}(0,l)S_F(l)\gamma_\mu S_F(l) V_{1\lambda}(-l,0)
%\right]_{\alpha\beta;\delta\gamma}
%G_{\lambda \nu}(l).
%\end{eqnarray}
each quantum correction becomes the same as that to the (pseudo) scalar
density and (axial) vector current operators
\begin{eqnarray}
I^{(c)}_{VA+AV} &=&
2T_S\left(S{C}^{-1}\circledast{C}P\right)
-2T_P\left(P{C}^{-1}\circledast{C}S\right)
\\
I^{(c)}_{VA-AV} &=&
 T_V\left( V{C}^{-1}\circledast{C}A\right)
+T_A\left( A{C}^{-1}\circledast{C}V\right),
\\
I^{(c')}_{VA+AV} &=&
 2T_P\left( SC^{-1}\circledast CP\right)
-2T_S\left( PC^{-1}\circledast CS\right)
\\
I^{(c')}_{VA-AV} &=&
 T_A\left( VC^{-1}\circledast CA\right)
+T_V\left( AC^{-1}\circledast CV\right).
\end{eqnarray}
where the direct product $\circledast$ means
\begin{eqnarray}
\left(\Gamma\circledast\Gamma'\right)_{\alpha\beta;\gamma\delta}
=\left(\Gamma\right)_{\alpha\gamma}\left(\Gamma'\right)_{\delta\beta}
\end{eqnarray}

Performing the Fierz rearrangement and the charge conjugation the vertex
function is transformed into the same spinor structure as at the tree
level without any mixing
\begin{eqnarray}
I^{(c)}_{VA+AV}+I^{(c')}_{VA+AV} &=&
-\left(T_S+T_P\right)\left(V \otimes A + A \otimes V\right),
\\
I^{(c)}_{VA-AV}+I^{(c')}_{VA-AV} &=&
-\left(T_V+T_A\right)\left( V \otimes A - A \otimes V \right).
\end{eqnarray}

\reseteqnum
% ------------------------ section 5 ------------------------
\section{Contribution from penguin diagrams}
\label{sec:penguin}

\subsection{One loop correction}

In order to evaluate contributions from the penguin diagram we consider 
a four-quark operator of the following form
\begin{eqnarray}
Q^{(k)}_{XY} &=&
\left(T^{(k)}\right)_{ab;cd}
\left(\Gamma_X\otimes\Gamma_Y\right)_{\alpha\beta;\gamma\delta}
\sum_{q=u,d,s}\alpha^{(n)}_q
\left(\bs_{a\alpha}d_{b\beta}\right)\left(\bq_{c\gamma}q_{d\delta}\right),
\end{eqnarray}
where $k=2n-1$ or $2n$ and coefficients $\alpha^{(n)}_q$ and $T^{(k)}$
are defined in \eqn{eqn:alpha-factor} and \eqn{eqn:color-factor}.
Then we evaluate the penguin diagram contribution to the amputated four
quark vertex function
\begin{eqnarray}
I_{k;XY}&=&
\vev{Q^{(k)}_{XY}
s_{a\alpha}(p_1)\bd_{b\beta}(p_2)q_{c\gamma}(p_3)\bq_{d\delta}(p_4)
}_{\rm 1PI},
\end{eqnarray}
which is given by the Feynman diagrams in Fig.~\ref{fig:penguin}.
All the external momentum are set to in-coming direction and the
internal gluon momentum is given by $p=p_1+p_2=-(p_3+p_4)$.
%The quark mass is kept non-zero at this stage and the on-shell condition
%$i\pslash+m=0$ may be set for the external quark momentum later.

The vertex correction is given in the form
\begin{eqnarray}
%I_{2;XY}&=&
%\left(-\frac{1}{2N}1\wt{\otimes}1+\frac{1}{2}1\wt{\odot}1\right)
%\left(\Gamma_X\otimes V_{1\mu}(p_3,p_4)\right)I^{P(1)}_{Y;\nu}(p,m_u)
%G_{\mu\nu}(p),
%\\
I_{2n-1;XY}&=&
J_{\rm pen}\Bigl(
\alpha_d^{(n)} I^{P(2)}_{XY;\mu}(p,m_d)\otimes V_{1\nu}(p_3,p_4)
+\alpha_s^{(n)} I^{P(2)}_{YX;\mu}(p,m_s)\otimes V_{1\nu}(p_3,p_4)
\Bigr)G_{\mu\nu}(p),
\nn\\&&
\label{eqn:I2n-1}
\\
I_{2n;XY}&=&
J_{\rm pen}\sum_{q=u,d,s}\alpha_q^{(n)} I^{P(1)}_{Y;\mu}(p,m_q)
\left(\Gamma_X \otimes V_{1\nu}(p_3,p_4)\right)G_{\mu\nu}(p),
\label{eqn:I2n}
\end{eqnarray}
where the color factor is given by
\begin{eqnarray}
J_{\rm pen}=\left(-\frac{1}{2N}1\wt{\otimes}1+\frac{1}{2}1\wt{\odot}1\right)
\end{eqnarray}
and we define the following loop integrals
\begin{eqnarray}
&&
I^{P(1)}_{Y;\mu}(p,m_q)=-\int_{-\pi}^\pi\frac{d^4l}{(2\pi)^4}
\tr\left(\Gamma_Y S_q(l-p) V_{1\mu}(-l+p,l) S_q(l) \right),
\label{eqn:penguin1}
\\&&
I^{P(2)}_{XY;\mu}(p,m_q)=\int_{-\pi}^\pi\frac{d^4l}{(2\pi)^4}
\left(\Gamma_X S_q(l-p) V_{1\mu}(-l+p,l)S_q(l)\Gamma_Y\right).
\label{eqn:penguin2}
\end{eqnarray}
We notice $I_{1;XY}=0$.
The contributions of $I^{P(1c)}_{Y;\mu}$ and $I^{P(2c)}_{XY;\mu}$ should
also be included where the gluon vertex in the loop is replaced with
$V^{(c)}_{1\mu}$ from the clover term.

We calculate the following loop integrals
\begin{eqnarray}
&&
I^{P(1)}_{V_\nu,\mu}(p,m_q)=-\tr\left(\gamma_\nu I^P_\mu(p,m_q) \right),
\quad
I^{P(1)}_{A_\nu,\mu}(p,m_q)=-\tr\left(\gamma_\nu\gamma_5 I^P_\mu(p,m_q) \right),
\label{eqn:IP1}
\\&&
I^{P(2)}_{VA;\mu}(p,m_q)=\gamma_\nu I^P_\mu(p,m_q)\gamma_\nu\gamma_5,
\quad
I^{P(2)}_{AV;\mu}(p,m_q)=\gamma_\nu\gamma_5 I^P_\mu(p,m_q) \gamma_\nu,
\label{eqn:IP2}
\\&&
I^{P}_{\mu}(p,m_q)=\int\frac{d^4l}{(2\pi)^4}S_F(l-p) V_{1\mu}(-l+p,l) S_F(l)
\label{eqn:Ipmu}
\end{eqnarray}
according to the standard procedure of lattice
perturbation theory \cite{Bernard:1987rw}, {\it i.e., }
we expand the functions in terms of the gluon momentum $p_\mu$ and the
quark mass $m$.
%\begin{eqnarray}
%I^{P(i)}_{\mu}(p,m)&=&
%\frac{1}{a^2}I^{P(i)}_{\mu}(0,0)
%\nn\\&&
%+\frac{1}{a}\left(
%m\left.\frac{\p}{\p(am)}I^{P(i)}_{\mu}(ap,am)\right|_{p=m=0}
%+p_\rho\left.\frac{\p}{\p(ap_\rho)}I^{P(i)}_{\mu}(ap,am)\right|_{p=m=0}
%\right)
%\nn\\&&
%+\Biggl(
%m^2\frac{1}{2}\left.\frac{\p^2}{\p (am)^2}I^{P(i)}_{\mu}(ap,am)\right|_{p=m=0}
%+p_\rho p_\lambda\frac{1}{2}
%\left.\frac{\p^2}{\p(ap_\rho)\p(ap_\lambda)}I^{P(i)}_{\mu}(ap,am)\right|_{p=m=0}
%\nn\\&&\quad
%+mp_\rho
%\left.\frac{\p^2}{\p(am)\p(ap_\rho)}I^{P(i)}_{\mu}(ap,am)\right|_{p=m=0}
%\Biggr)
%\nn\\&&
%+{\cal O}(a).
%\end{eqnarray}

The vertex functions satisfy the vector Ward-Takahashi
identity \cite{Bernard:1987rw}
\begin{eqnarray}
&&
\sum_{\mu}2\sin\frac{p_\mu}{2}I^{P}_{\mu}(p,m)=0,\quad
\label{eqn:WT-id1}
\\&&
\sum_{\mu}2\sin\frac{p_\mu}{2}I^{P(c)}_{\mu}(p,m)=0
\label{eqn:WT-id2}
\end{eqnarray}
arising from the identity
\begin{eqnarray}
&&
\sum_{\mu}2\sin\frac{p_\mu}{2}V_{1\mu}(-l+p,l)=S_F^{-1}(l-p)-S_F^{-1}(l),
\\&&
\sum_{\mu}2\sin\frac{p_\mu}{2}V^{(c)}_{1\mu}(-l+p,l)=0.
\end{eqnarray}
Hence the loop corrections should be proportional to
\begin{eqnarray}
\lim_{a\to0}I^{P}_\mu(p,m)
\propto\left(p^2\delta_{\mu\nu}-p_\mu p_\nu\right).
\label{eqn:allowed}
\end{eqnarray}
We notice that a term proportional to $\sigma_{\mu\nu}p_\nu$ is also
allowed by the identity.
However this term gives the same form of contribution as
\eqn{eqn:allowed} when expanded in terms of the external momentum and
substituted into \eqn{eqn:IP1} and \eqn{eqn:IP2}.
A detailed discussion shall be given later in Sec.~\ref{penguin-Oa}.

The only non-vanishing candidate is
\begin{eqnarray}
I^{P(i)}_{Y/XY;\mu}(p,m)&=&
\frac{1}{2}p_\alpha p_\beta
\left.\frac{\p^2}{\p(ap_\alpha)\p(ap_\beta)}
I^{P(i)}_{Y/XY;\mu}(p,m)\right|_{p=m=0}
+{\cal O}(a),
\end{eqnarray}
which has logarithmic divergence and should be regularized with some
infra-red regulator.
We shall adopt the gluon momentum $p_\mu$ as a regulator and the
regularization term is defined by the similar loop integral
\begin{eqnarray}
I^{P(i)}_{Y/XY;\mu}(p)_{\rm IR}&=&
\int_{-\pi}^{\pi} \frac{d^4 l}{(2 \pi)^4}\theta\left(\pi^2-l^2\right)
L^{P(i)}_{Y/XY;\mu}(l,p)_{\rm IR},
\end{eqnarray}
where $L^{P(i)}_{Y/XY;\mu}(l,p)_{\rm IR}$ is the same integrand given in
\eqn{eqn:IP1}, \eqn{eqn:IP2} but with all the Feynman rules replaced
with that in the continuum and the quark mass set to zero.

The loop integral is evaluated with a subtraction
\begin{eqnarray}
I^{P(i)}_{Y/XY;\mu}(p,m)&=&
I^{P(i)}_{Y/XY;\mu}(p,m)-I^{P(i)}_{Y/XY;\mu}(p)_{\rm IR}
+I^{P(i)}_{Y/XY;\mu}(p)_{\rm IR}.
\nn\\&=&
\frac{1}{2}p_\alpha p_\beta
\left.\frac{\p^2}{\p(ap_\alpha)\p(ap_\beta)}
\left(I^{P(i)}_{Y/XY;\mu}(p,m)
-I^{P(i)}_{Y/XY;\mu}(p)_{\rm IR}\right)\right|_{p=m=0}
+{\cal O}(a)
\nn\\&&
+I^{P(i)}_{Y/XY;\mu}(p)_{\rm IR}.
\end{eqnarray}
The first term is finite and can be evaluated numerically.
\begin{eqnarray}
\frac{\p^2}{\p(ap_\alpha)\p(ap_\beta)}
\left(I^{P(1)}_{V_\nu,\mu}(0)-I^{P(1)}_{V_\nu,\mu}(0)_{\rm IR}\right)
&=&
\frac{i}{16\pi^2}\Bigl(
\left(5.09290(43)\right)\delta_{\mu\nu}\delta_{\alpha\beta}
\nn\\&&
-\left(1.88003(27)\right)
\left(\delta_{\mu\alpha}\delta_{\nu\beta}+\delta_{\nu\alpha}\delta_{\mu\beta}
\right)\Bigr),
\\
\frac{\p^2}{\p(ap_\alpha)\p(ap_\beta)}
\left(I^{P(1)}_{A_\nu,\mu}(0)-I^{P(1)}_{A_\nu,\mu}(0)_{\rm IR}\right)
&=&0,
\\
\frac{\p^2}{\p(ap_\alpha)\p(ap_\beta)}
\left(I^{P(2)}_{VA;\mu}(0)-I^{P(2)}_{VA;\mu}(0)_{\rm IR}\right)
&=&
\frac{\p^2}{\p(ap_\alpha)\p(ap_\beta)}
\left(I^{P(2)}_{AV;\mu}(0)-I^{P(2)}_{AV;\mu}(0)_{\rm IR}\right)
\nn\\&=&
\frac{1}{2}
\frac{\p^2}{\p(ap_\alpha)\p(ap_\beta)}
\left(I^{P(1)}_{V_\nu,\mu}(0)-I^{P(1)}_{V_\nu,\mu}(0)_{\rm IR}\right)
\gamma_\nu\gamma_5.
\nn\\
\end{eqnarray}
The second term has a logarithmic divergence and is calculated
analytically
\begin{eqnarray}
&&
I^{P(1)}_{V_\nu,\mu}(p)_{\rm IR}=
\frac{4i}{16\pi^2}\left[
\left(-p^2\delta_{\mu\nu}+p_\mu p_\nu\right)
\frac{1}{3}\left(\ln\frac{\pi^2}{a^2p^2}+\frac{5}{6}\right)
-\frac{p^2}{6}\delta_{\mu\nu}+\frac{1}{2a^2}\delta_{\mu\nu}
\right],
\\&&
I^{P(1)}_{A_\nu,\mu}(p)_{\rm IR}=0,
\\&&
I^{P(2)}_{VA;\mu}(p)_{\rm IR}=
I^{P(2)}_{AV;\mu}(p)_{\rm IR}=
\frac{1}{2}I^{P(1)}_{V_\nu,\mu}(p)_{\rm IR}\gamma_\nu\gamma_5.
\end{eqnarray}
We shall drop the last $1/a^2$ divergent term since the corresponding
term is absent in the finite part since it is evaluated in terms of a
derivative with external momentum.

Contribution from the clover term is given by replacing the gluon
interaction vertex with $V^{(c)}_{1\mu}$ in \eqn{eqn:penguin1} and
\eqn{eqn:penguin2}.
The loop integral has no IR divergence and can be evaluated numerically.
\begin{eqnarray}
\frac{\p^2}{\p(ap_\alpha)\p(ap_\beta)}I^{P(1c)}_{V_\nu,\mu}(0)&=&
\frac{i}{16\pi^2}\Bigl(
-\left(2.90088(27)\right)
\delta_{\mu\nu}\delta_{\alpha\beta}
+\left(1.45031(13)\right)
\left(\delta_{\mu\alpha}\delta_{\nu\beta}+\delta_{\nu\alpha}\delta_{\mu\beta}
\right)\Bigr),
\nn\\
\\
\frac{\p^2}{\p(ap_\alpha)\p(ap_\beta)}I^{P(1c)}_{A_\nu,\mu}(0)&=&0,
\\
\frac{\p^2}{\p(ap_\alpha)\p(ap_\beta)}I^{P(2c)}_{VA;\mu}(0)&=&
\frac{\p^2}{\p(ap_\alpha)\p(ap_\beta)}I^{P(2c)}_{AV;\mu}(0)
=\frac{1}{2}
\frac{\p^2}{\p(ap_\alpha)\p(ap_\beta)}I^{P(1c)}_{V_\nu,\mu}(0)
\gamma_\nu\gamma_5.
\end{eqnarray}

\subsection{Tree level contribution}
\label{sec:tree-level}
\begin{figure}
\begin{center}
\begin{picture}(420,120)(0,0)
%\Text(50,120)[l]{Penguin diagram}
\Text(15,90)[l]{$s_a^\alpha$}
\Text(9,70)[l]{$-p$}
\Text(15,10)[l]{$\ovl{d}_b^\beta$}
\Text(15,30)[l]{$p$}
\Text(165,90)[l]{$\ovl{q}_d^\delta$}
\Text(165,10)[l]{$q_c^\gamma$}
\Text(100,80)[l]{$q$}
\Text(85,5)[l]{$l \rightarrow$}
\Text(80,100)[l]{$\leftarrow l$}
\Vertex(50,50){4}
\Vertex(60,50){4}
\ArrowLine(0,5)(50,50)
\ArrowLine(50,50)(0,95)
\CArc(100,50)(40,0,360)
\ArrowLine(160,50)(160,0)
\ArrowLine(160,100)(160,50)
\Text(240,90)[l]{$s_a^\alpha$}
\Text(240,10)[l]{$\ovl{d}_b^\beta$}
\Text(385,90)[l]{$\ovl{q}_d^\delta$}
\Text(385,10)[l]{$q_c^\gamma$}
\Text(320,80)[l]{$q$}
\Text(305,5)[l]{$l \rightarrow$}
\Text(300,100)[l]{$\leftarrow l$}
\Vertex(280,43){4}
\Vertex(280,57){4}
\ArrowLine(220,5)(280,43)
\ArrowLine(280,57)(220,95)
\CArc(320,50)(40,190,170)
\ArrowLine(380,50)(380,0)
\ArrowLine(380,100)(380,50)
\end{picture}
\end{center}
\caption{Tree level contribution to the $\Delta S=1$ four-quark
operator.
$\alpha,\beta$ and $a,b$ label Dirac and color indices respectively.
An external momentum $p$ is in-coming direction}
\label{fig:penguin-tree}
\end{figure}
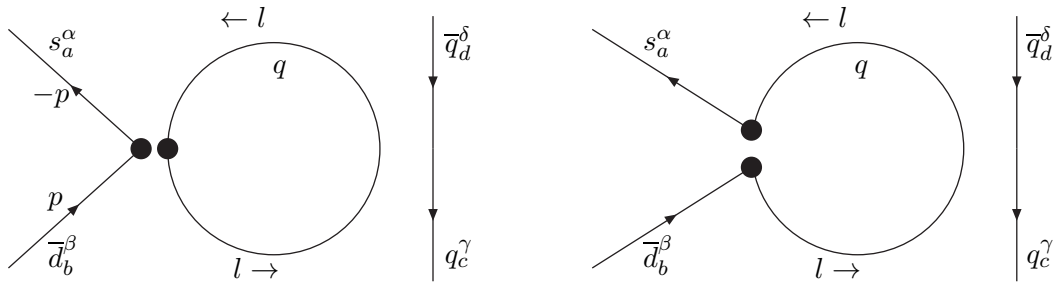

We consider the tree level contribution to the four quark operators given
in Fig.~\ref{fig:penguin-tree}.
These diagrams may give a power subtraction with lower dimensional
operators.
We shall evaluate the amputated quark bilinear vertex function given by
\begin{eqnarray}
I_{k;XY}^{\rm (sub)}&=&
\vev{Q^{(k)}_{XY} s_{a\alpha}(-p)\bd_{b\beta}(p)}_{\rm 1PI}.
\end{eqnarray}
For each operator we have the following vertex function
\begin{eqnarray}
I_{2n-1;XY}^{\rm (sub)}&=&
-N_c\delta_{ab}\left(\Gamma_X\right)_{\alpha\beta}
\sum_{q=u,d,s}\alpha^{(n)}_qI^{(2)}_{Y}(m_q)
\nn\\&&
+\alpha^{(n)}_d\delta_{ab}I^{\rm (sub)}_{XY}(m_d)_{\alpha\beta}
+\alpha^{(n)}_s\delta_{ab}I^{\rm (sub)}_{YX}(m_s)_{\alpha\beta},
\\
I_{2n;XY}^{\rm (sub)}&=&
-\delta_{ab}\left(\Gamma_X\right)_{\alpha\beta}
\sum_{q=u,d,s}\alpha^{(n)}_qI^{(2)}_{Y}(m_q)
\nn\\&&
+N_c\alpha^{(n)}_d\delta_{ab}I^{\rm (sub)}_{XY}(m_d)_{\alpha\beta}
+N_c\alpha^{(n)}_s\delta_{ab}I^{\rm (sub)}_{YX}(m_s)_{\alpha\beta},
\end{eqnarray}
where we have two kinds of loop integrals in the above
\begin{eqnarray}
&&
I^{\rm (sub)}_{XY}(m)=\int\frac{d^4l}{(2\pi)^4}\Gamma_XS_F(l,m)\Gamma_Y,
\\&&
I^{(2)}_{Y}(m)=\int\frac{d^4l}{(2\pi)^4}\tr\left(\Gamma_YS_F(l,m)\right)=0
\end{eqnarray}
The latter vanishes for both operator $Y=V, A$.
An explicit calculation gives
\begin{eqnarray}
&&
I^{\rm(sub)}_{VA}(m)=I^{\rm (sub)}(am)\gamma_5,
\\&&
I^{\rm(sub)}_{AV}(m)=-I^{\rm (sub)}(am)\gamma_5,
\\&&
I^{\rm (sub)}(am)=\int\frac{d^4l}{(2\pi)^4}
\frac{4W(l,am)}{\ovl{l}^2 + W(l,am)^2}.
\label{eqn:wilson-term}
\end{eqnarray}

Substituting this result we obtain
\begin{eqnarray}
&&
I_{2n-1;VA}^{\rm (sub)}=-I_{2n-1;AV}^{\rm (sub)}=
\alpha^{(n)}_d\delta_{ab}\left(\gamma_5\right)_{\alpha\beta}
\left(I^{\rm (sub)}(m_d)-I^{\rm (sub)}(m_s)\right),
\\&&
I_{2n;VA}^{\rm (sub)}=-I_{2n;AV}^{\rm (sub)}=
N\alpha^{(n)}_d\delta_{ab}\left(\gamma_5\right)_{\alpha\beta}
\left(I^{\rm (sub)}(m_d)-I^{\rm (sub)}(m_s)\right),
\end{eqnarray}
which may be evaluated with an expansion in the quark mass
\begin{eqnarray}
I^{\rm (sub)}(m)&=&
\frac{1}{a^2}m\frac{d}{d(am)}I^{\rm (sub)}(0)
+\frac{1}{a}m^2\frac{1}{2}\frac{d^2}{d(am)^2}I^{\rm (sub)}(0)
+m^3\frac{1}{6}\frac{d^3}{d(am)^3}I^{\rm (sub)}(0)
+{\cal O}(a).
\nn\\
\end{eqnarray}
The numerical evaluation gives
\begin{eqnarray}
&&
%I^{(1)}(0)=\frac{1}{16\pi^2}\left(148.23780(34)\right)
%\\&&
\frac{d}{d(am)}I^{\rm (sub)}(0)=\frac{1}{16\pi^2}\left(-21.46586(54)\right),
\\&&
\frac{d^2}{d(am)^2}I^{\rm (sub)}(0)
=\frac{1}{16\pi^2}\left(-14.9157(11)\right).
\label{eqn:tree-level}
\end{eqnarray}
It may not be a good idea to expand in terms of the quark mass since
these coefficients are rather large and furthermore
${d^3}/{d(am)^3}I^{\rm (sub)}(0)$ term has an infra red divergence at
$m=0$.

This contribution introduces a mixing with the lower dimensional
bilinear operator $(\bs\gamma_5d)$ multiplied with a mass difference
$(m_d-m_s)$ as is given in \eqn{eqn:pseudo-scalar}.
It is clear from \eqn{eqn:wilson-term} that the mixing is due to the
chiral symmetry breaking effect in the Wilson fermion.

\reseteqnum
% ------------------------ section 6 ------------------------
\section{Renormalization factor in $\ovl{\rm MS}$ scheme}
\label{sec:renormalization}

We renormalize the lattice bare operators $Q^{(k)}_{\rm lat}$ to obtain the
renormalized operator $Q^{(k)}_{\ovl{\rm MS}}$.
We adopt the $\ovl{\rm MS}$ scheme with DRED or NDR.
The renormalization of the operator is given by
\begin{eqnarray}
Q^{(i)}_{\ovl{\rm MS}}=Z_{ij}^gQ^{(j)}_{\rm lat}
+Z_i^{\rm pen}Q^{\rm pen}_{\rm lat}
+Z_i^{\rm sub}O^{\rm sub}_{\rm lat}
\end{eqnarray}
where $Q^{(j)}_{\rm lat}$ is the four quark operators on the lattice,
$Q_{\rm pen}^{\rm lat}$ is the QCD penguin operator and
$O_{\rm sub}^{\rm lat}$ is a lower dimensional operator to be
subtracted.
$Z_{ij}^g$ comes from the gluon exchanging diagrams.
$Z_i^{\rm pen}$ is the contribution from the penguin diagrams.

\subsection{Gluon exchanging diagrams}

For gluon exchanging diagram we sum up all the contributions from three
diagrams $(a)$, $(b)$, $(c)$ and multiply by the color factor.
Here we show its explicit form for the $\Delta S=1$ operators.
\begin{eqnarray}
I_{VA+AV}^{(k=1,3,9)}&=&
\left(
\frac{N^2-2}{2N}\left(T_V+T_A\right)+\frac{1}{2N}\left(T_S+T_P\right)
\right)
\left(1\wt{\otimes}1\right)\left(V \otimes A+A \otimes V\right)
\nn\\&&
+\frac{1}{2}\left(T_V+T_A-T_S-T_P\right)
\left(1\wt{\odot}1\right)\left(V \otimes A+A \otimes V\right)
%\nn\\&\to&
%\left(
%\frac{N^2-2}{2N}\left(T_V+T_A\right)+\frac{1}{2N}\left(T_S+T_P\right)
%\right)Q_{1,3,9}
%\nn\\&&
%+\frac{1}{2}\left(T_V+T_A-T_S-T_P\right)Q_{2,4,10}
\\
I_{VA+AV}^{(k=2,4,10)}&=&
\left(\frac{N^2-2}{2N}\left(T_V+T_A\right)
+\frac{1}{2N}\left(T_S+T_P\right)\right)
\left(1\wt{\odot}1\right)\left(V \otimes A+A \otimes V\right)
\nn\\&&
+\frac{1}{2}\left(T_V+T_A-T_S-T_P\right)
\left(1\wt{\otimes}1\right)\left(V \otimes A+A \otimes V\right)
%\nn\\&\to&
%\left(\frac{N^2-2}{2N}\left(T_V+T_A\right)
%+\frac{1}{2N}\left(T_S+T_P\right)\right)Q_{2,4,10}
%\nn\\&&
%+\frac{1}{2}\left(T_V+T_A-T_S-T_P\right)Q_{1,3,9}
\\
I_{VA-AV}^{(k=5,7)}&=&
\left(
\frac{N^2}{2N}\left(T_V+T_A\right)-\frac{1}{2N}\left(T_S+T_P\right)
\right)
\left(1\wt{\otimes}1\right)\left(V \otimes A-A \otimes V\right)
\nn\\&&
+\frac{1}{2}\left(-T_V-T_A+T_S+T_P\right)
\left(1\wt{\odot}1\right)\left(V \otimes A-A \otimes V\right)
%\nn\\&=&
%\left(\frac{N}{2}\left(T_V+T_A\right)-\frac{1}{2N}\left(T_S+T_P\right)\right)
%Q_{5,7}
%\nn\\&&
%+\frac{1}{2}\left(-T_V-T_A+T_S+T_P\right)Q_{6,8}
\\
I_{VA-AV}^{(k=6,8)}&=&
\left(\frac{N^2-1}{2N}\left(T_S+T_P\right)\right)
\left(1\wt{\odot}1\right)\left(V \otimes A-A \otimes V\right)
%\nn\\&\to&
%\left(\frac{N^2-1}{2N}\left(T_S+T_P\right)\right)Q_{6,8}
\end{eqnarray}
From these vertex functions one can easily see that the one loop
correction to the four quark operators is given in a form
\begin{eqnarray}
Q^{(i)}_{\rm one-loop}=T^{\rm lat}_{ij}Q^{(j)}_{\rm tree},
\end{eqnarray}
where $Q^{(j)}_{\rm tree}$ is a tree level operator.
The correction factors are given by
\begin{eqnarray}
T^{\rm lat}_{11}&=&
T^{\rm lat}_{22}=T^{\rm lat}_{33}=T^{\rm lat}_{44}
=T^{\rm lat}_{99}=T^{\rm lat}_{10,10}
=\frac{N^2-2}{2N}\left(T_V+T_A\right)+\frac{1}{2N}\left(T_S+T_P\right)
\nn\\&=&
\frac{g^2}{16\pi^2}
\left(-\frac{N^2+2}{N}\ln\left(\lambda a\right)^2
+\frac{N^2-2}{2N}\left(V_V+V_A\right)+\frac{1}{2N}\left(V_S+V_P\right)\right),
\\
T^{\rm lat}_{55}&=&
T^{\rm lat}_{77}
=\frac{N}{2}\left(T_V+T_A\right)-\frac{1}{2N}\left(T_S+T_P\right)
\nn\\&=&
\frac{g^2}{16\pi^2}
\left(-\frac{N^2-4}{N}\ln\left(\lambda a\right)^2
+\frac{N}{2}\left(V_V+V_A\right)
-\frac{1}{2N}\left(+V_S+V_P\right)
\right),
\\
T^{\rm lat}_{66}&=&
T^{\rm lat}_{88}
=\frac{N^2-1}{2N}\left(T_S+T_P\right)
\nn\\&=&
\frac{g^2}{16\pi^2}
\left(-4\frac{N^2-1}{N}\ln\left(\lambda a\right)^2
+\frac{N^2-1}{2N}\left(V_S+V_P\right)\right),
\\
T^{\rm lat}_{12}&=&
T^{\rm lat}_{21}=T^{\rm lat}_{34}=T^{\rm lat}_{43}
=T^{\rm lat}_{9,10}=T^{\rm lat}_{10,9}
=\frac{1}{2}\left(T_V+T_A-T_S-T_P\right)
\nn\\&=&
\frac{g^2}{16\pi^2}\frac{1}{2}
\left(6\ln\left(\lambda a\right)^2
+V_V+V_A-V_S-V_P\right),
\\
T^{\rm lat}_{56}&=&
T^{\rm lat}_{78}
=\frac{1}{2}\left(-T_V-T_A+T_S+T_P\right)
\nn\\&=&
\frac{g^2}{16\pi^2}\frac{1}{2}\left(
-6\ln\left(\lambda a\right)^2
-V_V-V_A+V_S+V_P\right),
\end{eqnarray}
where $\lambda$ is a gluon mass introduced for infrared
regularization.

The renormalization factor is given by taking a ratio of quantum
correction with that in the $\ovl{\rm MS}$ scheme multiplied with the
quark wave function renormalization factor $Z_2$
\begin{eqnarray}
&&
Z^g_{ii}(\mu a)=
\frac{\left(Z_2^{\ovl{\rm MS}}\right)^2\left(1+T_{ii}^{\ovl{\rm MS}}\right)}
{\left(Z_2^{\rm lat}\right)^2\left(1+T_{ii}^{\rm lat}\right)},
\\&&
Z^g_{ij}(\mu a)=T_{ij}^{\ovl{\rm MS}}-T_{ij}
\quad (i \neq j).
\end{eqnarray}
The correction factor in the DRED $\ovl{\rm MS}$ scheme is given by
\begin{eqnarray}
&&
T^{\ovl{\rm MS}}_{11}=T^{\ovl{\rm MS}}_{22}
=T^{\ovl{\rm MS}}_{33}=T^{\ovl{\rm MS}}_{44}
=T^{\ovl{\rm MS}}_{99}=T^{\ovl{\rm MS}}_{10,10}
=\left(\frac{N^2+2}{N}\right)V^{\ovl{\rm MS}},
\\&&
T^{\ovl{\rm MS}}_{12}=T^{\ovl{\rm MS}}_{21}
=T^{\ovl{\rm MS}}_{34}=T^{\ovl{\rm MS}}_{43}
=T^{\ovl{\rm MS}}_{9,10}=T^{\ovl{\rm MS}}_{10,9}
=-3V^{\ovl{\rm MS}},
\\&&
T^{\ovl{\rm MS}}_{55}=T^{\ovl{\rm MS}}_{77}
=\left(\frac{N^2-4}{N}\right)V^{\ovl{\rm MS}},
\\&&
T^{\ovl{\rm MS}}_{56}=T^{\ovl{\rm MS}}_{78}=3V^{\ovl{\rm MS}},
\\&&
T^{\ovl{\rm MS}}_{66}=T^{\ovl{\rm MS}}_{88}
=4\frac{N^2-1}{N}V^{\ovl{\rm MS}},
\\&&
V^{\ovl{\rm MS}} = \frac{g^2}{16\pi^2}
\left(\log\left(\frac{\mu^2}{\lambda^2}\right)+1\right).
\end{eqnarray}
The quark wave function renormalization factor is given by
Ref.~\cite{Aoki:1998ar} and the result in the Feynman gauge is
\begin{equation}
\left(\frac{Z_2^{\ovl{\rm MS}}}{Z_2^{\rm lat}}\right)(\mu a)
= 1+\frac{g^2}{16\pi^2}C_F\left(
-\log (\mu a)^2 +\Sigma_1^{\overline{\rm MS}} -\Sigma_1
\right),
\end{equation}
where 
\begin{eqnarray}
&&
%\Sigma_1^{\overline{\rm MS}}({\rm NDR})=\frac{1}{2},\quad
\Sigma_1^{\overline{\rm MS}}({\rm DRED})=-\frac{1}{2},
\\&&
\Sigma_1=\Sigma_1^{(0)}+c_{\rm SW}\Sigma_1^{(1)}+c_{\rm SW}^2\Sigma_1^{(2)}.
\end{eqnarray}
The numerical value of $\Sigma_1^{(n)}$ is given in table \ref{tab:self}.

Substituting the above results we have
\begin{eqnarray}
Z^g_{11}(\mu a)&=&
Z^g_{22}(\mu a)=Z^g_{33}(\mu a)=Z^g_{44}(\mu a)=Z^g_{99}(\mu a)
=Z^g_{10,10}(\mu a)
\nn\\&=&
1+\frac{g^2}{16\pi^2}\left(\frac{3}{N}\ln\left(\mu a\right)^2+z^g_{11}\right),
\\
Z^g_{55}(\mu a)&=&Z^g_{77}(\mu a)
=1+\frac{g^2}{16\pi^2}\left(-\frac{3}{N}\ln\left(\mu a\right)^2+z^g_{55}\right),
\\
Z^g_{66}(\mu a)&=&Z^g_{88}(\mu a)
=1+\frac{g^2}{16\pi^2}\left(\frac{3\left(N^2-1\right)}{N}\ln\left(\mu a\right)^2
+z^g_{66}\right),
\\
Z^g_{12}(\mu a)&=&
Z^g_{21}(\mu a)=Z^g_{34}(\mu a)=Z^g_{43}(\mu a)=Z^g_{9,10}(\mu a)
=Z^g_{10,9}(\mu a)
\nn\\&=&
\frac{g^2}{16\pi^2}\left(-3\ln\left(\mu a\right)^2+z^g_{12}\right),
\\
Z^g_{56}(\mu a)&=&Z^g_{78}(\mu a)
=\frac{g^2}{16\pi^2}\left(3\ln\left(\mu a\right)^2+z^g_{56}\right),
\\
Z^g_{65}(\mu a)&=&Z^g_{87}(\mu a)=\frac{g^2}{16\pi^2}z^g_{65}=0.
\\
z^g_{11}&=&
\frac{N^2+2}{N}
-\frac{N^2-2}{2N}\left(V_V+V_A\right)-\frac{1}{2N}\left(V_S+V_P\right)
+2C_F\left(\Sigma_1^{\overline{\rm MS}}-\Sigma_1\right),
\\
z^g_{55}&=&
\frac{N^2-4}{N}
-\frac{N}{2}\left(V_V+V_A\right)+\frac{1}{2N}\left(V_S+V_P\right)
+2C_F\left(\Sigma_1^{\overline{\rm MS}}-\Sigma_1\right),
\\
z^g_{66}&=&
4\frac{N^2-1}{N}-\frac{N^2-1}{2N}\left(V_S+V_P\right)
+2C_F\left(\Sigma_1^{\overline{\rm MS}}-\Sigma_1\right),
\\
z^g_{12}&=&
-3-\frac{1}{2}\left(V_V+V_A-V_S-V_P\right),
\\
z^g_{56}&=&-z^g_{12}.
\end{eqnarray}
The numerical result is given in table \ref{tab:gluon-exchanging} as an
expansion in $c_{\rm SW}$
\begin{eqnarray}
z^g_{ij}=z^{g(0)}_{ij}+c_{\rm SW}z^{g(1)}_{ij}+c_{\rm SW}^2z^{g(2)}_{ij}
\end{eqnarray}
for $N=3$.

We need to subtract the evanescent operators $E^{(i)}$ in the
$\ovl{{\rm MS}}$ scheme, which comes from the difference of dimensionality 
from four for gamma matrices in the operator vertex
\cite{Buras:1989xd,Herrlich:1994kh}.
The evanescent operators in the DRED scheme is given by
\begin{eqnarray}
&&
E^{(i)}=E^{\rm (DRED)}_{ij}E_j,
\\&&
E_{1,3,9}=
\left(1\wt{\otimes}1\right)
\left(\frac{4}{n}\ovl{\gamma}_\nu^L \otimes \ovl{\gamma}_\nu^L
-\gamma_\nu^L \otimes \gamma_\nu^L\right)\frac{2}{\epsilon},
\\&&
E_{2,4,10}=\left(1\wt{\odot}1\right)
\left(\frac{4}{n}\ovl{\gamma}_\nu^L \otimes \ovl{\gamma}_\nu^L
-\gamma_\nu^L \otimes \gamma_\nu^L\right)\frac{2}{\epsilon},
\\&&
E_{5,7}=\left(1\wt{\otimes}1\right)
\left(\frac{4}{n}\ovl{\gamma}_\nu^L \otimes \ovl{\gamma}_\nu^R
-\gamma_\nu^L \otimes \gamma_\nu^R\right)\frac{2}{\epsilon},
\\&&
E_{6,8}=\left(1\wt{\odot}1\right)
\left(\frac{4}{n}\ovl{\gamma}_\nu^L \otimes \ovl{\gamma}_\nu^R
-\gamma_\nu^L \otimes \gamma_\nu^R\right)\frac{2}{\epsilon},
\\&&
E^{\rm (DRED)}_{11}=E^{\rm (DRED)}_{33}=E^{\rm (DRED)}_{99}
=-\frac{g^2}{16\pi^2}N,
\\&&
E^{\rm (DRED)}_{22}=E^{\rm (DRED)}_{44}=E^{\rm (DRED)}_{10,10}
=\frac{g^2}{16\pi^2}N,
\\&&
E^{\rm (DRED)}_{12}=E^{\rm (DRED)}_{34}=E^{\rm (DRED)}_{9,10}
=\frac{g^2}{16\pi^2},
\\&&
E^{\rm (DRED)}_{21}=E^{\rm (DRED)}_{43}=E^{\rm (DRED)}_{10,9}
=-\frac{g^2}{16\pi^2},
\\&&
E^{\rm (DRED)}_{55}=E^{\rm (DRED)}_{77}=-\frac{g^2}{16\pi^2}2C_F,
\\&&
E^{\rm (DRED)}_{66}=E^{\rm (DRED)}_{88}=\frac{g^2}{16\pi^2}\frac{1}{N},
\\&&
E^{\rm (DRED)}_{65}=E^{\rm (DRED)}_{87}=-\frac{g^2}{16\pi^2},
\end{eqnarray}
where $n$ is the dimension of the loop momentum.
$\ovl{\gamma}_\nu^{L/R}$ and $\gamma_\nu^{L/R}$ are $n=(4-\epsilon)$ and
four dimensional gamma  matrix with the chiral projection
$(1\mp\gamma_5)$.

The conversion formula to the NDR scheme is as follows
\cite{Aoki:2000ee}
\begin{eqnarray}
&&
\left(z^g_{11}\right)^{\rm NDR}
=\left(z^g_{11}\right)^{\rm DRED}-\frac{N^2-6}{2N},
\\&&
\left(z^g_{55}\right)^{\rm NDR}
=\left(z^g_{55}\right)^{\rm DRED}-\frac{N^2-8}{2N},
\\&&
\left(z^g_{66}\right)^{\rm NDR}
=\left(z^g_{66}\right)^{\rm DRED}-\frac{N^2-4}{N},
\\&&
\left(z^g_{12}\right)^{\rm NDR}=\left(z^g_{12}\right)^{\rm DRED}-\frac{5}{2},
\\&&
\left(z^g_{56}\right)^{\rm NDR}=\left(z^g_{56}\right)^{\rm DRED}-\frac{7}{2},
\\&&
\left(z^g_{65}\right)^{\rm NDR}=-3
\end{eqnarray}
with corresponding evanescent operators.
The numerical value of $z^g_{ij}$ is given in table
\ref{tab:gluon-exchanging-NDR} for NDR.

\subsection{Penguin diagrams}

Taking a summation of the finite part and the IR divergent term the one
loop correction from the penguin diagram is given by
\begin{eqnarray}
&&
%I^{\rm pen}_{1;VA}=I^{\rm pen}_{1;AV}=0,
%\\&&
%I^{\rm pen}_{2;VA}=0,
%\\&&
%I^{\rm pen}_{2;AV}=
%T^{\rm lat}_{\rm pen}(p)\left(-\frac{1}{N}1\wt{\otimes}1+1\wt{\odot}1\right)
%\left(\gamma_{\mu}\gamma_5\otimes\gamma_{\mu}\right),
%\\&&
I^{\rm pen}_{2n-1;VA}=
\alpha_d^{(n)}T^{\rm lat}_{\rm pen}(p)
\left(-\frac{1}{N}1\wt{\otimes}1+1\wt{\odot}1\right)
\left(\gamma_{\mu}\gamma_5\otimes\gamma_{\mu}\right),
\\&&
I^{\rm pen}_{2n-1;AV}=
\alpha_d^{(n)}T^{\rm lat}_{\rm pen}(p)
\left(-\frac{1}{N}1\wt{\otimes}1+1\wt{\odot}1\right)
\left(\gamma_{\mu}\gamma_5\otimes\gamma_{\mu}\right),
\\&&
I^{\rm pen}_{2n;VA}=0,
\\&&
I^{\rm pen}_{2n;AV}=
\left(\sum_{q=u,d,s}\alpha_q^{(n)}\right)T^{\rm lat}_{\rm pen}(p)
\left(-\frac{1}{N}1\wt{\otimes}1+1\wt{\odot}1\right)
\left(\gamma_\mu\gamma_5 \otimes \gamma_{\mu}\right),
\\&&
T^{\rm lat}_{\rm pen}(p)
=\frac{g^2}{16\pi^2}\frac{2}{3}\left(\ln{a^2p^2}+V_{\rm pen}^{\rm lat}\right),
\end{eqnarray}
where we adopt the on-shell condition for external quarks
\begin{eqnarray}
-i\pslash_3+m_q=0,\quad
i\pslash_4+m_q=0,\quad
p=-p_3-p_4.
\end{eqnarray}
The finite part is expanded as
\begin{eqnarray}
V_{\rm pen}^{\rm lat}=V_{\rm pen}^{(0)}+c_{\rm SW}V_{\rm pen}^{(1)}
\label{eqn:vpen-lat}
\end{eqnarray}
with coefficients given in table \ref{tab:penguin-lat}.

We notice that the above vertex corresponds to a four fermi operator of
the form
\begin{eqnarray}
&&
\left(\bs_a\gamma_\mu\gamma_5d_a\right)
\sum_{q=u,d,s}\left(\bq_b\gamma_\mu q_b\right)
,\quad
\left(\bs_a\gamma_\mu\gamma_5d_b\right)
\sum_{q=u,d,s}\left(\bq_b\gamma_\mu q_a\right)
\end{eqnarray}
and is given by a linear combination of $Q^{(3)}$, $Q^{(4)}$, $Q^{(5)}$,
$Q^{(6)}$, which defines the penguin operator
\begin{eqnarray}
Q^{\rm pen}=\left(Q^{(4)}_{VA+AV}+Q^{(6)}_{VA-AV}\right)
-\frac{1}{N}\left(Q^{(3)}_{VA+AV}+Q^{(5)}_{VA-AV}\right).
\end{eqnarray}
The one loop correction from the penguin diagram to the four quark
operators is written as
\begin{eqnarray}
Q^{(i)}_{\rm one-loop}=\left(T^{\rm pen}_{i}\right)^{\rm lat}
Q^{\rm pen}_{\rm tree},
\end{eqnarray}
where $Q^{\rm pen}_{\rm tree}$ is the penguin operator at tree level.
The correction factor is given by
\begin{eqnarray}
&&
\left(T^{\rm pen}_{i}\right)^{\rm lat}=\frac{g^2}{16\pi^2}\frac{C(Q^{(i)})}{3}
\left(\ln{a^2p^2}+V_{\rm pen}^{\rm lat}\right)
\end{eqnarray}
with operator dependent factor
\begin{eqnarray}
&&
C\left(Q^{(1)}\right)=0,
\\&&
C\left(Q^{(2)}\right)=1,
\\&&
C\left(Q^{(3)}\right)=2,
\\&&
C\left(Q^{(4)}\right)=C\left(Q^{(6)}\right)=\sum_{q=u,d,s}\alpha_q^{(2)}=N_f,
\\&&
C\left(Q^{(5)}\right)=C\left(Q^{(7)}\right)=0,
\\&&
C\left(Q^{(8)}\right)=C\left(Q^{(10)}\right)=\sum_{q=u,d,s}\alpha_q^{(4)}
=N_u-\frac{N_d}{2},
\\&&
C\left(Q^{(9)}\right)=-1.
\end{eqnarray}

The correction factor in the $\ovl{\rm MS}$ scheme is given in a similar
form
\begin{eqnarray}
&&
Q^{(i)}_{\rm one-loop}=\left(T^{\rm pen}_{i}\right)^{\ovl{\rm MS}}
Q^{\rm pen}_{\rm tree},
\\&&
\left(T^{\rm pen}_{i}\right)^{\ovl{\rm MS}}
=\frac{g^2}{16\pi^2}\frac{C(Q^{(i)})}{3}
\left(\ln\left(\frac{p^2}{\mu^2}\right)-\frac{5}{3}-c\left(Q^{(i)}\right)
\right),
\end{eqnarray}
where the scheme dependent finite term is given by
\begin{eqnarray}
&&
c^{\rm (NDR)}\left(Q^{(2)}\right)=c^{\rm (NDR)}\left(Q^{(2n-1)}\right)=-1,
\quad
c^{\rm (NDR)}\left(Q^{(2n)}\right)=0,
\\&&
c^{\rm (DRED)}\left(Q^{(2)}\right)=c^{\rm (DRED)}\left(Q^{(2n-1)}\right)
=c^{\rm (DRED)}\left(Q^{(2n)}\right)=\frac{1}{4}
\end{eqnarray}
for $n\ge2$.

Combining these two contributions the renormalization factor for the
penguin operator is given by
\begin{eqnarray}
&&
Z_i^{\rm pen}=
\left(T^{\rm pen}_{i}\right)^{\ovl{\rm MS}}
-\left(T^{\rm pen}_{i}\right)^{\rm lat}
=\frac{g^2}{16\pi^{2}}\frac{C(Q^{(i)})}{3}\left(-\ln a^2\mu^2
+z_i^{\rm pen}\right),
\\&&
z_i^{\rm pen}=-V_{\rm pen}^{\rm lat}-\frac{5}{3}-c(Q^{(i)}).
\end{eqnarray}
Numerical value of the finite part is given in table \ref{tab:penguin}.

\subsection{Subtraction of lower dimensional operator}

As was discussed in Sec.~\ref{sec:tree-level} the lower dimensional
operator
\begin{eqnarray}
O^{\rm sub}_{\rm lat}=\bs\gamma_5d
\end{eqnarray}
mixes with the four quark operators.
The subtraction factor is given by
\begin{eqnarray}
&&
Z_{2n-1}^{\rm (sub)}=
-2\alpha^{(n)}_d\left(I^{\rm (sub)}(m_d)-I^{\rm (sub)}(m_s)\right),
\quad(n=3,4),
\\&&
Z_{2n}^{\rm (sub)}=
-2N\alpha^{(n)}_d\left(I^{\rm (sub)}(m_d)-I^{\rm (sub)}(m_s)\right),
\quad(n=3,4),
\\&&
Z_{2n-1}^{\rm (sub)}=Z_{2n}^{\rm (sub)}=0,\quad
(n=1,2,5).
\end{eqnarray}
We may be better to evaluate these factors nonperturbatively for
numerical simulation.

\subsection{Mean field improvement}

The mean field improvement is given by subtracting the tadpole
contribution in the renormalization factor and replacing it by a
nonperturbative value $u$ given in terms of the average plaquette
$u=P^{1/4}$ for example.
The tadpole contribution resides only in $\Sigma_1$ of the quark wave
function renormalization factor $Z_2$.
The mean field improvement works for the diagonal renormalization factor
$Z^g_{ii}$ from the gluon exchanging diagram.
In the improved renormalization we shall use the renormalization factor
\begin{eqnarray}
u^2Z^{g({\rm MF})}_{ii}
\end{eqnarray}
instead of $Z^g_{ii}$.
$Z^{g({\rm MF})}_{ii}$ is given by replacing the finite term $z^g_{ii}$
by $z^{g({\rm MF})}_{ii}$ in which the tadpole contribution is
subtracted.
The $c_{\rm SW}$ dependent part is not affected by the mean field
improvement.
The numerical value is given in table \ref{tab:mena-field}.

\reseteqnum
% ------------------------ section 7 ------------------------
\section{$O(a)$ improvement coefficients}
\label{sec:oa}

In order for the on-shell $O(a)$ improvement program to work one
need to adopt the rotated field for the operator
\begin{eqnarray}
&&
\psi_c = \left[ 1-\frac{ar}{2}
 \left( z \gamma_\mu \rightD_\mu - (1-z) m \right) \right] \psi,
\\&&
\bpsi_c = \bpsi \left[ 1-\frac{ar}{2}
 \left( -z \gamma_\mu \leftD_\mu - (1-z) m \right) \right]
\end{eqnarray}
in addition to the improvement of the action.
We shall set the on-shell condition
\begin{eqnarray}
\left(\gamma_\mu \rightD_\mu + m_q\right) \psi_q =0,\quad
\bpsi_q\left(-\gamma_\mu \leftD_\mu+m_q\right)=0
\end{eqnarray}
for the quark fields and adopt $z=0$ for simplicity.
The bare mass $m_q$ is defined by subtracting the additive mass
correction from the bare Wilson fermion mass.

A typical form of the tree level improved four fermi operator is given
by
\begin{eqnarray}
{\cal O}_{\rm lat}&=&%{\Gamma\Gamma'}
\left(1+\frac{ar}{2}\left(m_1+m_2+m_3+m_4\right)\right)
\left(\bpsi_1\Gamma\psi_2\right)\left(\bpsi_3\Gamma'\psi_4\right)
+{\cal O}(a^2),
\label{eqn:Oabare-operator}
\end{eqnarray}
which we shall adopt for our lattice bare operator.
Each quark fields $\psi_i$ has an incoming external momentum $p_i$.
In the following we set $r=1$.

The renormalization relation is given by
\begin{eqnarray}
Q^{(i)}_{\ovl{\rm MS}}&=&
Z_{ij}^gQ^{(j)}_{\rm lat}
+Z_i^{\rm pen}Q^{\rm pen}_{\rm lat}
+Z_i^{\rm sub}O^{\rm sub}_{\rm lat}
\nn\\&&
-g^2aB_{ij}Q^{(j)}_{\rm lat}
-g^2aB'_{in}O_{n,{\rm lat}}
-g^2aB^{\rm pen}_qQ^{\rm pen}_{\rm lat}
-g^2aC_{ij}\wt{Q}^{(j)}_{\rm lat}
\label{eqn:msbar-oa}
\end{eqnarray}
with $O(a)$ subtractions, where $Q_{\rm lat}$ is a tree level improved
lattice bare operator for the $K\to\pi\pi$ decay.
$O_{n,{\rm lat}}$ represents four fermi operators with wrong chirality
given in \eqn{eqn:SP} - \eqn{eqn:TT}.
$B$'s are proportional to the quark mass $m_q$.
$\wt{Q}$'s are dimension seven operators proportional to the quark
external momentum $p_\mu$.

\subsection{Contribution from gluon exchanging diagrams}

We shall evaluate the $O(a)$ correction for massive quarks in this
subsection.
The correction has already been calculated in
Ref.~\cite{Constantinou:2010zs} for gluon exchanging diagrams with
massless quarks.

For the gluon exchanging correction all the ten operators $Q^{(i)}$ are
not distinguishable but we have only four distinction
$O^{(k)}_{VA\pm AV}$, where $k$ takes even or odd for the color factor.
So we shall evaluate the one loop correction to the following four
operators
\begin{eqnarray}
&&
O_1=O^{(o)}_{VA+AV},\quad
O_2=O^{(e)}_{VA+AV},
\\&&
O_3=O^{(o)}_{VA-AV},\quad
O_4=O^{(e)}_{VA-AV},
\end{eqnarray}
for which we shall need six more operators to mix with at $O(g^2a)$
\begin{eqnarray}
&&
O_5=O^{(o)}_{SP-PS},\quad
O_6=O^{(e)}_{SP-PS},
\label{eqn:SP}
\\&&
O_7=O^{(o)}_{SP+PS},\quad
O_8=O^{(e)}_{SP+PS},
\label{eqn:PS}
\\&&
O_9=O^{(o)}_{\wt{T}T},\quad
O_{10}=O^{(e)}_{\wt{T}T}.
\label{eqn:TT}
\end{eqnarray}
The flavor structure shall take the form given in \eqn{eqn:SPPS} -
\eqn{eqn:TTtilde} for a practical use in $K\to\pi\pi$ decay.

We consider the following four fermi operator
\begin{eqnarray}
&&
O_n=
\left(\Gamma_n\right)_{a\alpha b\beta;c\gamma d\delta}
\left(\bpsi_{1;a,\alpha}\psi_{2;b,\beta}\right)
\left(\bpsi_{3;c,\gamma}\psi_{4;d,\delta}\right),
\\&&
\left(\Gamma^{(k)}_{VA\pm AV}\right)_{a\alpha b\beta;c\gamma d\delta}=
\left(T^{(k)}\right)_{ab;cd}
\left(\mp V\otimes A-A\otimes V\right)_{\alpha\beta;\gamma\delta},
\\&&
\left(\Gamma^{(k)}_{SP\pm PS}\right)_{a\alpha b\beta;c\gamma d\delta}=
\left(T^{(k)}\right)_{ab;cd}
\left(S\otimes P\pm P\otimes S\right)_{\alpha\beta;\gamma\delta},
\\&&
\left(\Gamma^{(k)}_{\wt{T}T}\right)_{a\alpha b\beta;c\gamma d\delta}=
\left(T^{(k)}\right)_{ab;cd}
\left(\wt{T}\otimes T\right)_{\alpha\beta;\gamma\delta},
\end{eqnarray}
%\begin{eqnarray}
%&&
%O^{(k)}_{VA+AV}=
%\left(T^{(k)}\right)_{ab;cd}
%\left(-V\otimes A-A\otimes V\right)_{\alpha\beta;\gamma\delta}
%\left(\bpsi_{1;a,\alpha}\psi_{2;b,\beta}\right)
%\left(\bpsi_{3;c,\gamma}\psi_{4;d,\delta}\right),
%\\&&
%O^{(k)}_{VA-AV}=
%\left(T^{(k)}\right)_{ab;cd}
%\left(V\otimes A-A\otimes V\right)_{\alpha\beta;\gamma\delta}
%\left(\bpsi_{1;a,\alpha}\psi_{2;b,\beta}\right)
%\left(\bpsi_{3;c,\gamma}\psi_{4;d,\delta}\right),
%\\&&
%O^{(k)}_{SP-PS}=
%\left(T^{(k)}\right)_{ab;cd}
%\left(S\otimes P-P\otimes S\right)_{\alpha\beta;\gamma\delta}
%\left(\bpsi_{1;a,\alpha}\psi_{2;b,\beta}\right)
%\left(\bpsi_{3;c,\gamma}\psi_{4;d,\delta}\right),
%\\&&
%O^{(k)}_{SP+PS}=
%\left(T^{(k)}\right)_{ab;cd}
%\left(S\otimes P+P\otimes S\right)_{\alpha\beta;\gamma\delta}
%\left(\bpsi_{1;a,\alpha}\psi_{2;b,\beta}\right)
%\left(\bpsi_{3;c,\gamma}\psi_{4;d,\delta}\right),
%\\&&
%O^{(k)}_{\wt{T}T}=
%\left(T^{(k)}\right)_{ab;cd}
%\left(\wt{T}\otimes T\right)_{\alpha\beta;\gamma\delta}
%\left(\bpsi_{1;a,\alpha}\psi_{2;b,\beta}\right)
%\left(\bpsi_{3;c,\gamma}\psi_{4;d,\delta}\right),
%\end{eqnarray}
where $T^{(k)}$ represents the color factor.

The one loop contribution is written as
\begin{eqnarray}
I_{k;VA\pm AV}^{(a,b,c)}=J_k^{(a,b,c)}
\left(1+\frac{1}{2}a\left(m_{1}+m_2+m_{3}+m_4\right)\right)
I_{VA\pm AV}^{(a,b,c)}
\end{eqnarray}
including the $O(g^2a)$ terms, where $J_k^{(a,b,c)}$ is a color
factor given in Sec.~\ref{sec:gluon}.

As was mentioned there the one loop correction is given in terms of that
to the bilinear operator \eqn{eqn:current-correction}
\begin{eqnarray}
I_{VA\pm AV}^{(a+a')}&=&
\mp\left(G_V^{(12)} \otimes A\right)+\left(V \otimes G_A^{(34)}\right)
-\left(\left(G_A^{(12)} \otimes V\right)
+\left(A \otimes G_V^{(34)}\right)\right),
\label{eqn:Ia}
\nn\\
\\
I_{VA+AV}^{(b+b')}&=&
\left(G_V^{(14)} \odot A+V \odot G_A^{(23)}
+G_A^{(14)} \odot V+A \odot G_V^{(23)}\right),
\\
I_{VA-AV}^{(b+b')}&=&
-2\left( G_S^{(14)} \odot P+S \odot G_P^{(23)}
-G_P^{(14)} \odot S-P \odot G_S^{(23)}\right),
\\
I_{VA+AV}^{(c+c')}&=&
-2\biggl(
 \left(G_S^{(13)}{C}^{-1}\circledast{C}P\right)
+\left(SC^{-1}\circledast CG_P^{(42)}\right)
\nn\\&&
-\left(G_P^{(13)}{C}^{-1}\circledast{C}S\right)
-\left(PC^{-1}\circledast CG_S^{(42)}\right)\biggr),
\\
I_{VA-AV}^{(c+c')}&=&
 \left( G_V^{(13)}C^{-1}\circledast CA\right)
+\left( VC^{-1}\circledast CG_A^{(42)}\right)
\nn\\&&
+\left( G_A^{(13)}C^{-1}\circledast CV\right)
+\left( AC^{-1}\circledast CG_V^{(42)}\right),
\label{eqn:Ic}
\end{eqnarray}
where $G_\Gamma^{(ij)}$ is a one loop correction to the bilinear vertex
$\Gamma$ with i-th and j-th flavor contributes for the internal quark
line
\begin{eqnarray}
&&
G_\Gamma^{(ij)}=\int_{-\pi}^{\pi}\frac{d^4l}{(2\pi)^4}
V_{1\mu}(p_i,l-p_i)S_F(l-p_i,m_i)\Gamma
 S_F(l+p_j,m_j)V_{1\nu}(-l-p_j,p_j)G_{\mu\nu}(l).
\nn\\
\end{eqnarray}
The quark mass and external momentum is kept non-vanishing here.

The vertex correction can be expanded in terms of the quark mass and the
external momentum up to $O(a)$ according to Ref.~\cite{Taniguchi:1998pf}
\begin{eqnarray}
G_\Gamma^{(ij)}=T_\Gamma\Gamma
+T_\Gamma^{(m)}\frac{1}{2}a\left(m_i+m_j\right)\Gamma
+T_\Gamma^{(p)}ia\left(p_i+p_j\right)_\mu\wt{\Gamma}_\mu,
\label{eqn:bilinear-oa}
\end{eqnarray}
where the vertex $\wt{\Gamma}_\mu$ for operator subtraction is given as
$\wt{\Gamma}_\mu^+$ in Ref.~\cite{Taniguchi:1998pf}.
We notice that $O(g^2a\log a)$ term cancels by adopting the tree level
improvement condition $c_{\rm SW}=1$ and the on-shell condition
\eqn{eqn:on-shell} for the external quarks.
All the coefficients are proportional to $g^2$.

Performing an explicit evaluation with $c_{\rm SW}=1$ the $O(g^2a)$
coefficients are given by $T_{A}^{(p)}=-C_{A}$ and $T_{V}^{(p)}=C_{V}$,
where $C_{A/V}$ is defined in Ref.~\cite{Taniguchi:1998pf}.
We notice that definition of the tensor operator
$\sigma_{\mu\nu}=\frac{1}{2}\left[\gamma_\mu,\gamma_\nu\right]$
is different from that in Ref.~\cite{Taniguchi:1998pf}.
$T_P^{(p)}=T_S^{(p)}=0$ for on-shell quarks.
$T_\Gamma^{(m)}$ is equivalent to $V_\Gamma^{(1)}$ in
Ref.~\cite{Taniguchi:1998pf} and its explicit value can be reconstructed
by
\begin{eqnarray}
V_\Gamma^{(1)}=-B_\Gamma+b_0
\end{eqnarray}
in the reference.
We give numerical values in tables \ref{tab:bilinear-oap},
\ref{tab:bilinear-oam} for the Iwasaki gauge action in order to avoid
confusion.
The coefficients for the other gauge action can be reconstructed
according to the example from the numerical tables in
Ref.~\cite{Taniguchi:1998pf}.

Substituting into \eqn{eqn:Ia} - \eqn{eqn:Ic} the one loop correction is
expanded as
\begin{eqnarray}
I_{VA\pm AV}^{({\rm diagram})}=
I_{VA\pm AV}^{({\rm diagram})(g^2)}
+I_{VA\pm AV}^{({\rm diagram})(g^2am)}
+I_{VA\pm AV}^{({\rm diagram})(g^2ap)}.
\label{eqn:I-expansion}
\end{eqnarray}
Each term represents $O(g^2)$, $O(g^2am)$ and $O(g^2ap)$.

In this subsection we adopt the lattice scheme \eqn{eqn:lattice-scheme}
with gluon mass regulator implicitly and evaluate the full vertex
correction $G_1,\cdots,G_4$ for operators $O_{n=1\sim4}$.
For example $G_1$ is given by
\begin{eqnarray}
G_1&=&
\left(1+\frac{1}{2}a\left(m_{1}+m_2+m_{3}+m_4\right)\right)
\left(
\Gamma_{VA+AV}^{(o)}
+\sum_{{\rm diagram}=a,b,c}J_o^{({\rm diagram})}I_{VA+AV}^{({\rm diagram})}
\right).
\nn\\
\end{eqnarray}
with the tree level vertex $\Gamma_{VA+AV}^{(o)}$

According to Ref.~\cite{Taniguchi:1998pf} we rewrite the bare quark mass
in terms of the renormalized one
\begin{eqnarray}
&&
m_q=Z_mm_{qR},
\\&&
Z_m=1 + g^2 Z_m^{(1)},
\\&&
Z_m^{(1)}=C_F \left( -3\,L +z_m \right),\quad
L=-\frac{1}{16\pi^2}\ln\lambda^2a^2.
\end{eqnarray}
We then multiply the four quark vertex correction
with the wave function renormalization factor
$Z_{\psi_1}^{1/2}Z_{\psi_2}^{1/2}Z_{\psi_3}^{1/2}Z_{\psi_4}^{1/2}$
\begin{eqnarray}
Z_{\psi_q}^{1/2}&=&
1+\frac{1}{2}g^2Z_\psi^{(1)}+\frac{1}{2}a m_{qR} \left( -1 + g^2Z_\psi^{(a)}
\right),
\label{eqn:zpsi}
\\
Z_\psi^{(1)}&=& C_F \left( -L + \Sigma_1 \right),
\\
Z_\psi^{(a)}&=&C_F\left(\frac{7}{2}L+\frac{1}{2}\Sigma_1-z_m
+\Sigma_1^{(1)}\right).
\end{eqnarray}
Here $\Sigma_1$, $\Sigma_1^{(1)}$ and $z_m$ are constants introduced in 
Ref.~\cite{Taniguchi:1998pf} and is given in table \ref{tab:quark-oa}
for $c_{\rm SW}=1$.

The $O(a)$ and $O(g^2a\log a)$ terms are canceled in
\begin{eqnarray}
Z_{\psi_1}^{1/2}Z_{\psi_2}^{1/2}Z_{\psi_3}^{1/2}Z_{\psi_4}^{1/2}G_i
=Z_{ij}^{g(\rm lat)}\Gamma_j
+g^2aB_{ij}\Gamma_j
+g^2aC_{ij}\wt{\Gamma}_j
\label{eqn:coeff-BC}
\end{eqnarray}
and one can extract the renormalization factor $Z_{ij}^{g(\rm lat)}$ and
the $O(a)$ coefficient $B_{ij}$, $B'_{in}$, $C_{ij}$.
The renormalization factor $Z_{ij}^{g(\rm lat)}$ has already been used
in Sec.~\ref{sec:renormalization} to get that in the $\ovl{\rm MS}$
scheme.
Although an explicit form of the $O(a)$ coefficient is given in the
appendix \ref{sec:appendixA},
we mention that the lattice bare operator $Q^{(j)}_{\rm lat}$ is given
by $O_{n=1\sim4}$ and $O_{n,{\rm lat}}$ is given by $O_{n=5\sim10}$.
The dimension seven operator $\wt{Q}^{(j)}_{\rm lat}$ is given by
\begin{eqnarray}
\wt{Q}^{(j)}=\left(\wt{\Gamma}_j\right)_{a\alpha b\beta;c\gamma d\delta}
\left(\bpsi_{1;a,\alpha}\psi_{2;b,\beta}\right)
\left(\bpsi_{3;c,\gamma}\psi_{4;d,\delta}\right)
\end{eqnarray}
in terms of the vertex given in \eqn{eqn:gamma-tilde-1} -
\eqn{eqn:gamma-tilde-8}.

The same $O(a)$ coefficients appear in the renormalization relation
\eqn{eqn:msbar-oa} for the $\ovl{\rm MS}$ scheme.
These coefficients are written in terms of the one loop corrections
$T_\Gamma^{(m)}$, $T_\Gamma^{(p)}$, $\Sigma_1+\Sigma_1^{(1)}$ multiplied
with quark masses and external momentum as will be given in the
appendix.

\subsection{Contribution from penguin diagrams}
\label{penguin-Oa}
The one loop correction from the penguin diagram to the improved operator
\eqn{eqn:Oabare-operator} is given by a slight modification of the one
loop vertex \eqn{eqn:I2n-1} and \eqn{eqn:I2n} multiplied with the tree
level improvement factor
\begin{eqnarray}
\left(1+\frac{a}{2}\left(m_d+m_s\right)\right)\left(1+am_q\right)I_{2n-1/2n;XY}.
\end{eqnarray}
However this factor shall be canceled with the tree level contribution
of the wave function renormalization factor \eqn{eqn:zpsi} and we
abbreviate it.
We shall evaluate \eqn{eqn:IP1}, \eqn{eqn:IP2} by an expansion
in the quark mass and the external momentum.

Before performing the expansion we make use of the Ward-Takahashi
identity \eqn{eqn:WT-id1}, \eqn{eqn:WT-id2}.
We notice that the identity is valid on the lattice with a non-vanishing
quark mass and external momentum and impose a restriction on the one
loop correction \eqn{eqn:Ipmu}
\begin{eqnarray}
&&
I^{P}_{\mu}(p,m_q)=
\left(\hat{p}^2\delta_{\mu\lambda}-\hat{p}_\mu\hat{p}_\lambda\right)
l^{P}_\lambda(p,m_q)
+\sigma_{\mu\lambda}\hat{p}_\lambda\wt{l}^{P}(p,m_q),
\label{eqn:allowed-oa}
\\&&
\hat{p}=2\sin\frac{p}{2}.
\end{eqnarray}

Taking into account the Lorentz covariance each term is expanded to give
the $O(a)$ contribution
\begin{eqnarray}
&&
l^{P}_\lambda(p,m_q)=
l^{P}\gamma_\lambda
+m_ql^{P}_m\gamma_\lambda
+p_\lambda l^{P}_{pS}
+p_\alpha\sigma_{\alpha\lambda} l^{P}_{pA},
\\&&
\wt{l}^{P}(p,m_q)=
\wt{l}^{P}
+m_q\wt{l}^{P}_m
+m_q^2\wt{l}^{P}_{mm}
+p_\alpha\gamma_\alpha\wt{l}^{P}_{p}
+p^2\wt{l}^{P}_{pp}.
\end{eqnarray}
We substitute this expansion into a typical correction term
\eqn{eqn:allowed-oa} and substitute further into the one loop penguin
contributions \eqn{eqn:IP1} and \eqn{eqn:IP2}.
After a short algebraic calculation we find that only three terms
contribute to the one loop penguin diagram
\begin{eqnarray}
&&
I^{P(1)}_{V_\nu,\mu}(p,m_q)=
-4\left(\delta_{\mu\nu}p^2-p_\mu p_\nu\right)
\left(l^{P}+\wt{l}^{P}_{p}+am_ql^{P}_m\right),
\\&&
I^{P(1)}_{A_\nu,\mu}(p,m_q)=0,
\\&&
I^{P(2)}_{VA;\mu}(p,m_q)=I^{P(2)}_{AV;\mu}(p,m_q)
=\frac{1}{2}I^{P(1)}_{V_\nu,\mu}(p,m_q)\gamma_\nu\gamma_5.
\end{eqnarray}

From the clover term contribution
\begin{eqnarray}
I^{P(c)}_{\mu}(p,m_q)=\int\frac{d^4l}{(2\pi)^4}S_F(l-p,m_q)
 V_{1\mu}^{(c)}(-l+p,l) S_F(l,m_q)
\end{eqnarray}
we have the similar form of correction
\begin{eqnarray}
&&
I^{P(1c)}_{V_\nu,\mu}(p,m_q)=
-4\left(\delta_{\mu\nu}p^2-p_\mu p_\nu\right)
c_{\rm SW}\left(l^{P(c)}+\wt{l}^{P(c)}_{p}+am_ql^{P(c)}_m\right),
\\&&
I^{P(1)}_{A_\nu,\mu}(p,m_q)=0,
\\&&
I^{P(2c)}_{VA;\mu}(p,m_q)=I^{P(2c)}_{AV;\mu}(p,m_q)
=\frac{1}{2}I^{P(1c)}_{V_\nu,\mu}(p,m_q)\gamma_\nu\gamma_5.
\end{eqnarray}

The $O(g)$ terms have already been evaluated in Sec.~\ref{sec:penguin}
and we put the same result for the notation used here
\begin{eqnarray}
&&
l^P_1=\left(l^{P}+\wt{l}^{P}_{p}\right)
=\frac{ig}{16\pi^2}\frac{1}{3}\left(-\ln{a^2p^2}+1.7128269(84)\right),
\\&&
l^{P(c)}_1=\left(l^{P(c)}+\wt{l}^{P(c)}_{p}\right)
=\frac{ig}{16\pi^2}\frac{1}{3}\left(1.087821(3)\right).
\end{eqnarray}
We notice that both the $O(gam_q)$ coefficient $l^{P}_m$ and
$l^{P(c)}_m$ has a logarithmic IR divergence.
The same regularization scheme is also used here as was adopted in
Sec.~\ref{sec:penguin} and the coefficient is given by
\begin{eqnarray}
&&
l^{P}_{m}
=\frac{ig}{16\pi^2}\frac{1}{3}\left(\frac{5}{2}\ln{a^2p^2}-3.59121(36)\right),
%-1.19707(12)
\\&&
l^{P(c)}_{m}
=\frac{ig}{16\pi^2}\frac{1}{3}\left(-\frac{3}{2}\ln{a^2p^2}+0.74846(23)\right).
%+0.249487(78)
\end{eqnarray}

We substitute these results into the penguin contribution
\eqn{eqn:I2n-1} and \eqn{eqn:I2n} including all the contributions up to
$O(g^2a)$
\begin{eqnarray}
I_{2n-1;VA}&=&I_{2n-1;AV}
\nn\\&=&
J_{\rm pen}\sum_{q=d,s}\alpha_q^{(n)}
\left(I^{P(2)}_{VA;\mu}(p,m_q)+I^{P(2c)}_{VA;\mu}(p,m_q)\right)
\otimes
\left(V_{1\nu}(p_3,p_4)+V_{1\nu}^{(c)}(p_3,p_4)\right)
\nn\\&&\quad\times
G_{\mu\nu}(p)
\nn\\&=&
2igJ_{\rm pen}\sum_{q=d,s}\alpha_q^{(n)}
\Biggl(
\left(l^{P}_1+c_{\rm SW}l^{P(c)}_1\right)\gamma_\mu\gamma_5\otimes\gamma_\mu
\nn\\&&\quad
+m_q\left(l^{P}_m+c_{\rm SW}\left(l^{P}_1+l^{P(c)}_m\right)
+c_{\rm SW}^2l^{P(c)}_1\right)
\gamma_\mu\gamma_5\otimes\gamma_\mu
\nn\\&&\quad
+\left(1-c_{\rm SW}\right)\left(l^{P}_1+c_{\rm SW}l^{P(c)}_1\right)
\gamma_\mu\gamma_5\otimes\frac{i}{2}\left(p_3-p_4\right)_\mu
\Biggr),
\\
I_{2n;VA}&=&0,
\\
I_{2n;AV}&=&
J_{\rm pen}\sum_{q=u,d,s}\alpha_q^{(n)}
 \left(I^{P(1)}_{V_\mu;\nu}(p,m_q)+I^{P(1c)}_{V_\mu;\nu}(p,m_q)\right)
\nn\\&&\times
\left(\gamma_\mu\gamma_5 \otimes V_{1\rho}(p_3,p_4)
+\gamma_\mu\gamma_5 \otimes V_{1\rho}^{(c)}(p_3,p_4)\right)
G_{\nu\rho}(p)
\nn\\&=&
4igJ_{\rm pen}\sum_{q=u,d,s}\alpha_q^{(n)}
\Biggl(
\left(l^{P}_1+c_{\rm SW}l^{P(c)}_1\right)\gamma_\mu\gamma_5 \otimes \gamma_\mu
\nn\\&&\quad
+m_q\left(l^{P}_m+c_{\rm SW}\left(l^{P}_1+l^{P(c)}_m\right)
+c_{\rm SW}^2l^{P(c)}_1\right)
\gamma_\mu\gamma_5 \otimes \gamma_\mu
\nn\\&&\quad
+\left(1-c_{\rm SW}\right)
\left(l^{P}_1+c_{\rm SW}l^{P(c)}_1\right)
\left(\gamma_\mu\gamma_5 \otimes\frac{i}{2}\left(p_3-p_4\right)_\mu\right)
\Biggr).
\end{eqnarray}
Here we made use of on-shell conditions for the external momentum
\begin{eqnarray}
&&
%-i\pslash_1+m_1=0,\quad
%i\pslash_2+m_2=0
%\\&&
%-i\pslash_3+m_3=0,\quad
%i\pslash_4+m_4=0
%\\&&
i\left(\pslash_1+\pslash_2\right)=0,\quad
i\left(\pslash_3+\pslash_4\right)=0,
%-i\left(p_3+p_4\right)_\mu\gamma_\mu=-m_q+m_q=0
\\&&
(p_3-p_4)_\mu p_\mu=0,
%-\left(p_3-p_4\right)_\mu\left(p_3+p_4\right)_\mu=-p_3^2+p_4^2=m_q^2-m_q^2=0
\\&&
i\sigma_{\mu\nu}\left(p_3+p_4\right)_\nu
=i\left(p_3-p_4\right)_\mu
-2m_q\gamma_\mu.
\end{eqnarray}

We notice that the mixing with an operator
$\gamma_\mu\gamma_5\otimes i\left(p_3-p_4\right)_\mu$ drops if we set
the improvement coefficient $c_{\rm SW}=1$.
The $O(g^2a\log a)$ term in $l^{P}_{m}$ and $l^{P(c)}_{m}$ cancels in
a combination $l^{P}_m+c_{\rm SW}\left(l^{P}_1+l^{P(c)}_m\right)$ for
$c_{\rm SW}=1$.
The $O(g^2a)$ improvement is accomplished just by shifting 
$V_{\rm pen}^{\rm lat}$ in \eqn{eqn:vpen-lat} as
\begin{eqnarray}
V_{\rm pen}^{\rm lat}\to V_{\rm pen}^{\rm lat}+m_q\left(0.04210(43)\right).
\end{eqnarray}

\section{Conclusion}
\label{sec:conclusion}
In this paper we have calculated the one-loop contributions
to the renormalization factors for the parity odd four quark operators,
which contribute to the $K\to\pi\pi$ decay amplitude,
for the improved Wilson fermion action with the clover term and the
Iwasaki gauge action.
The operators are multiplicatively renormalizable without any mixing
with wrong operators that have different chiral structures except for
the lower dimensional operator.
The $O(g^2a)$ improvement coefficients are also calculated for massive
quarks imposing $c_{\rm SW}=1$ and the on-shell condition.

\section*{Acknowledgment}
I would like to thank K.~-I.~Ishikawa, N.~Ishizuka, A.~Ukawa and
T.~Yoshi\'e for valuable discussions.
This work is supported in part by Grants-in-Aid of the Ministry of
Education (Nos. 22540265, 23105701).

\appendix
\def\thesection{\Alph{section}}

\reseteqnum the one loop correction is
% ------------------------ appendix A ------------------------
\section{$O(a)$ contribution from gluon exchanging diagram}
\label{sec:appendixA}

In this appendix we evaluate the one loop correction from the gluon
exchanging diagrams with non-vanishing quark mass and the external
momentum.

We start from the one loop correction \eqn{eqn:Ia}-\eqn{eqn:Ic} and 
substitute into \eqn{eqn:bilinear-oa}.
The correction is expanded as \eqn{eqn:I-expansion}.
Explicit form of the $O(g^2am)$ terms are given by
\begin{eqnarray}
I_{VA\pm AV}^{(a)(g^2am)}&=&
\frac{1}{2}\left(\left(m_1+m_2\right)T_V^{(m)}+\left(m_3+m_4\right)T_A^{(m)}
\right)\left(V \otimes A\right)
\nn\\&&
\pm\frac{1}{2}\left(\left(m_1+m_2\right)T_A^{(m)}+\left(m_3+m_4\right)T_V^{(m)}
\right)\left(A \otimes V\right),
\\
I_{VA+AV}^{(b)(g^2am)}&=&
-\frac{1}{2}\left(
\left(m_1+m_4\right)T_V^{(m)}
+\left(m_2+m_3\right)T_A^{(m)}\right)\left(V \odot A\right)
\nn\\&&
-\frac{1}{2}\left(\left(m_1+m_4\right)T_A^{(m)}
+\left(m_2+m_3\right)T_V^{(m)}\right)\left(A \odot V\right),
\\
I_{VA-AV}^{(b)(g^2am)}&=&
-\left(\left(m_1+m_4\right)T_S^{(m)}+\left(m_2+m_3\right)T_P^{(m)}\right)
\left(S \odot P\right)
\nn\\&&
+\left(\left(m_1+m_4\right)T_P^{(m)}+\left(m_2+m_3\right)T_S^{(m)}\right)
\left(P \odot S\right),
\\
I_{VA+AV}^{(c)(g^2am)}&=&
\left(\left(m_1+m_3\right)T_S^{(m)}+\left(m_2+m_4\right)T_P^{(m)}\right)
\left(S{C}^{-1}\circledast{C}P\right)
\nn\\&&
-\left(\left(m_1+m_3\right)T_P^{(m)}+\left(m_2+m_4\right)T_S^{(m)}\right)
\left(P{C}^{-1}\circledast{C}S\right),
\\
I_{VA-AV}^{(c)(g^2am)}&=&
\frac{1}{2}\left(\left(m_1+m_3\right)T_V^{(m)}
+\left(m_2+m_4\right)T_A^{(m)}\right) \left( VC^{-1}\circledast CA\right)
\nn\\&&
+\frac{1}{2}\left(\left(m_1+m_3\right)T_A^{(m)}
+\left(m_2+m_4\right)T_V^{(m)}\right)\left( AC^{-1}\circledast CV\right).
\end{eqnarray}
Using the Fierz transformation with or without the charge conjugation we
rearrange the spinor structure
\begin{eqnarray}
&&
\left(V \odot A\right)
=-\frac{1}{2}\left(V\otimes A+A\otimes V\right)-S\otimes P+P\otimes S,
\\&&
\left(A \odot V\right)
=-\frac{1}{2}\left(V\otimes A+A\otimes V\right)+S\otimes P-P\otimes S,
\\&&
\left(S \odot P\right)
=\frac{1}{4}\left(S\otimes P+P\otimes S-V\otimes A+A\otimes V
 -\wt{T}\otimes T\right),
\\&&
\left(P \odot S\right)
=\frac{1}{4}\left(S\otimes P+P\otimes S+V\otimes A-A\otimes V
 -\wt{T}\otimes T\right),
\\&&
\left(S{C}^{-1}\circledast{C}P\right)
=\frac{1}{4}\left(S\otimes P+P\otimes S-V\otimes A-A\otimes V
 +\wt{T}\otimes T\right),
\\&&
\left(P{C}^{-1}\circledast{C}S\right)
=\frac{1}{4}\left(S\otimes P+P\otimes S+V\otimes A+A\otimes V
 +\wt{T}\otimes T\right),
\\&&
\left( VC^{-1}\circledast CA\right)
=-\frac{1}{2}\left(V\otimes A-A\otimes V\right)-S\otimes P+P\otimes S,
\\&&
\left( AC^{-1}\circledast CV\right)
=-\frac{1}{2}\left(V\otimes A-A\otimes V\right)+S\otimes P-P\otimes S.
\end{eqnarray}
Here we notice there appears an operator mixing with wrong chirality
proportional to the quark mass difference and chiral symmetry breaking
effect $(T_V-T_A)$ or $(T_S-T_P)$.

The $O(g^2ap)$ terms are given by
\begin{eqnarray}
I_{VA\pm AV}^{(a)(g^2ap)}&=&
T_A^{(p)}ia\left(p_3+p_4\right)_\mu\left(\gamma_\mu \otimes \gamma_5\right)
\pm T_A^{(p)}ia\left(p_1+p_2\right)_\mu\left(\gamma_5 \otimes \gamma_\mu\right)
\nn\\&&
+T_V^{(p)}ia\left(p_1+p_2\right)_\nu
\left(\sigma_{\mu\nu} \otimes \gamma_\mu\gamma_5\right)
\pm T_V^{(p)}ia\left(p_3+p_4\right)_\nu
\left(\gamma_\mu\gamma_5\otimes\sigma_{\mu\nu}\right),
\\
I_{VA+AV}^{(b)(g^2ap)}&=&
-T_A^{(p)}ia\left(p_2+p_3\right)_\mu\left(\gamma_\mu\odot\gamma_5\right)
-T_A^{(p)}ia\left(p_1+p_4\right)_\mu\left(\gamma_5\odot\gamma_\mu\right)
\nn\\&&
-T_V^{(p)}ia\left(p_1+p_4\right)_\nu
\left(\sigma_{\mu\nu}\odot\gamma_\mu\gamma_5\right)
-T_V^{(p)}ia\left(p_2+p_3\right)_\nu
\left(\gamma_\mu\gamma_5\odot\sigma_{\mu\nu}\right),
\\
I_{VA-AV}^{(b)(g^2ap)}&=&
-2T_S^{(p)}ia\left(p_1+p_4\right)_\mu\left(\gamma_\mu\odot\gamma_5\right)
-2T_P^{(p)}ia\left(p_2+p_3\right)_\mu\left(1\odot\gamma_\mu\gamma_5\right)
\nn\\&&
+2T_P^{(p)}ia\left(p_1+p_4\right)_\mu\left(\gamma_\mu\gamma_5\odot1\right)
+2T_S^{(p)}ia\left(p_2+p_3\right)_\mu\left(\gamma_5\odot\gamma_\mu\right),
\\
I_{VA+AV}^{(c)(g^2ap)}&=&
 2T_S^{(p)}ia\left(p_1+p_3\right)_\mu
\left(\gamma_\mu{C}^{-1}\circledast{C}\gamma_5\right)
+2T_P^{(p)}ia\left(p_2+p_4\right)_\mu
\left(1C^{-1}\circledast C\gamma_\mu\gamma_5\right)
\nn\\&&
-2T_P^{(p)}ia\left(p_1+p_3\right)_\mu
\left(\gamma_\mu\gamma_5{C}^{-1}\circledast{C}1\right)
-2T_S^{(p)}ia\left(p_2+p_4\right)_\mu
\left(\gamma_5C^{-1}\circledast C\gamma_\mu\right),
\nn\\
\\
I_{VA-AV}^{(c)(g^2ap)}&=&
T_A^{(p)}ia\left(p_2+p_4\right)_\mu
\left( \gamma_\mu C^{-1}\circledast C\gamma_5\right)
+T_A^{(p)}ia\left(p_1+p_3\right)_\mu
\left( \gamma_5C^{-1}\circledast C\gamma_\mu\right)
\nn\\&&
+T_V^{(p)}ia\left(p_1+p_3\right)_\nu
\left( \sigma_{\mu\nu}C^{-1}\circledast C\gamma_\mu\gamma_5\right)
+T_V^{(p)}ia\left(p_2+p_4\right)_\nu
\left( \gamma_\mu\gamma_5C^{-1}\circledast C\sigma_{\mu\nu}\right).
\nn\\
\end{eqnarray}
Using the momentum conservation relation $p_1+p_2+p_3+p_4=0$ and the
on-shell condition
\begin{eqnarray}
&&
\left(i\pslash_i+m_i\right)\psi_i(p_i)=0,\quad
\bpsi_i(p_i)\left(-i\pslash_i+m_i\right)=0,
\label{eqn:on-shell}
\\&&
\left(i\pslash_i+m_i\right)C^{-1}\bpsi^T_i(p_i)=0,\quad
\psi^T_i(p_i)C\left(-i\pslash_i+m_i\right)=0
\end{eqnarray}
we rewrite the correction
\begin{eqnarray}
I_{VA\pm AV}^{(a)(g^2ap)}&=&
-T_A^{(p)}a\left(m_1-m_2\right)\left(S \otimes P\right)
\mp T_A^{(p)}a\left(m_3-m_4\right)\left(P \otimes S\right)
\nn\\&&
+T_V^{(p)}ia\left(p_1+p_2\right)_\nu
\left(\sigma_{\mu\nu} \otimes \gamma_\mu\gamma_5\right)
\pm T_V^{(p)}ia\left(p_3+p_4\right)_\nu
\left(\gamma_\mu\gamma_5\otimes\sigma_{\mu\nu}\right),
\\
I_{VA+AV}^{(b)(g^2ap)}&=&
 T_A^{(p)}a\left(m_1-m_4\right)\left(S\odot P\right)
+T_A^{(p)}a\left(-m_2+m_3\right)\left(P\odot S\right),
\nn\\&&
-T_V^{(p)}ia\left(p_1+p_4\right)_\nu
\left(\sigma_{\mu\nu}\odot\gamma_\mu\gamma_5\right)
-T_V^{(p)}ia\left(p_2+p_3\right)_\nu
\left(\gamma_\mu\gamma_5\odot\sigma_{\mu\nu}\right),
\\
I_{VA-AV}^{(b)(g^2ap)}&=&
-2\left(T_S^{(p)}a\left(m_1-m_4\right)+T_P^{(p)}a\left(m_2+m_3\right)\right)
\left(S\odot P\right)
\nn\\&&
+2\left(T_P^{(p)}a\left(m_1+m_4\right)+T_S^{(p)}a\left(-m_2+m_3\right)\right)
\left(P\odot S\right)
\\
I_{VA+AV}^{(c)(g^2ap)}&=&
 2\left(T_S^{(p)}a\left(m_1-m_3\right)+T_P^{(p)}a\left(m_2+m_4\right)\right)
\left(SC^{-1}\circledast CP\right)
\nn\\&&
-2\left(T_P^{(p)}a\left(m_1+m_3\right)+T_S^{(p)}a\left(-m_2+m_4\right)\right)
\left(PC^{-1}\circledast CS\right),
\\
I_{VA-AV}^{(c)(g^2ap)}&=&
-T_A^{(p)}a\left(m_1-m_3\right)
\left(SC^{-1}\circledast CP\right)
-T_A^{(p)}a\left(-m_2+m_4\right)
\left(PC^{-1}\circledast CS\right)
\nn\\&&
+T_V^{(p)}ia\left(p_1+p_3\right)_\nu
\left( \sigma_{\mu\nu}C^{-1}\circledast C\gamma_\mu\gamma_5\right)
+T_V^{(p)}ia\left(p_2+p_4\right)_\nu
\left( \gamma_\mu\gamma_5C^{-1}\circledast C\sigma_{\mu\nu}\right).
\nn\\
\end{eqnarray}

Taking summation of all the contributions for $VA+AV$ and $VA-AV$ and
multiplying the wave function renormalization factor we extract the
$O(g^2am)$ coefficients in the \label{eqn:coeff-BC}
\begin{eqnarray}
B_{11}&=&
g^2aM\left(
C_F\left(\Sigma_1+\Sigma_1^{(1)}\right)
+\left(C_F-\frac{1}{2N}\right)\left(T_V^{(m)}+T_A^{(m)}\right)
+\frac{1}{2N}\left(T_S^{(m)}+T_P^{(m)}\right)
\right)
\nn\\&&
+g^2\frac{1}{N}\left(T_S^{(p)}am_{(14)}+T_P^{(p)}aM\right),
\\
B_{12}&=&
g^2aM\frac{1}{2}\left(T_V^{(m)}+T_A^{(m)}-T_S^{(m)}-T_P^{(m)}\right)
-g^2\left(T_S^{(p)}am_{(14)}+T_P^{(p)}aM\right),
\\
B_{13}&=&
-g^2am_{(12)}\left(
C_F\left(T_V^{(m)}-T_A^{(m)}\right)
+\frac{1}{2N}T_A^{(p)}\right),
\\
B_{14}&=&
g^2am_{(12)}\frac{1}{2}T_A^{(p)},
\\
B'_{15}&=&
g^2am_{14}\left(
-\frac{1}{2N}\left(T_V^{(m)}-T_A^{(m)}\right)
+2C_F T_A^{(p)}\right),
\\
B'_{16}&=&
g^2am_{14}\frac{1}{2}\left(T_V^{(m)}-T_A^{(m)}\right),
\\
B'_{17}&=&
g^2\Biggl(
am_{(13)}\frac{1}{2N}\left(T_S^{(m)}-T_P^{(m)}\right)
\nn\\&&\quad
+\left(2C_F+\frac{1}{2N}\right)T_A^{(p)}am_{(13)}
+\frac{1}{N}\left(T_S^{(p)}am_{(12)}-T_P^{(p)}am_{(13)}\right)
\Biggr),
\\
B'_{18}&=&
-g^2\left(am_{(13)}\frac{1}{2}\left(T_S^{(m)}-T_P^{(m)}\right)
+\frac{1}{2}T_A^{(p)}am_{(13)}
+\left(T_S^{(p)}am_{(12)}-T_P^{(p)}am_{(13)}\right)\right),
\\
B'_{19}&=&
g^2\left(
am_{(13)}\frac{1}{2N}\left(T_S^{(m)}-T_P^{(m)}\right)
-\frac{1}{2N}T_A^{(p)}am_{(13)}
+\frac{1}{N}\left(T_S^{(p)}am_{(12)}-T_P^{(p)}am_{(13)}\right)\right),
\nn\\&&
\\
B'_{1,10}&=&
g^2\left(-am_{(13)}\frac{1}{2}\left(T_S^{(m)}-T_P^{(m)}\right)
+\frac{1}{2}T_A^{(p)}am_{(13)}
-\left(T_S^{(p)}am_{(12)}-T_P^{(p)}am_{(13)}\right)\right),
\\
B_{21}&=&
g^2aM\frac{1}{2}\left(T_V^{(m)}+T_A^{(m)}-T_S^{(m)}-T_P^{(m)}\right)
-g^2\left(T_S^{(p)}am_{(14)}+T_P^{(p)}aM\right),
\\
B_{22}&=&
g^2aM\left(
C_F\left(\Sigma_1+\Sigma_1^{(1)}\right)
+\left(C_F-\frac{1}{2N}\right)\left(T_V^{(m)}+T_A^{(m)}\right)
+\frac{1}{2N}\left(T_S^{(m)}+T_P^{(m)}\right)\right)
\nn\\&&
+g^2\frac{1}{N}\left(T_S^{(p)}am_{(14)}+T_P^{(p)}aM\right),
\\
B_{23}&=&
-\frac{1}{2}g^2m_{(12)}\left(T_V^{(m)}-T_A^{(m)}\right),
\\
B_{24}&=&
\frac{1}{2N}g^2m_{(12)}\left(T_V^{(m)}-T_A^{(m)}\right)
+g^2am_{(12)}C_FT_A^{(p)},
\\
B'_{25}&=&g^2am_{(14)}T_A^{(p)},
\\
B'_{26}&=&
g^2C_Fm_{14}\left(T_V^{(m)}-T_A^{(m)}\right)
-g^2am_{(14)}\frac{1}{N}T_A^{(p)},
\\
B'_{27}&=&
-g^2am_{(13)}\frac{1}{2}\left(T_S^{(m)}-T_P^{(m)}\right)
+g^2\left(T_A^{(p)}am_{(13)}-T_S^{(p)}am_{(12)}+T_P^{(p)}am_{(13)}\right),
\\
B'_{28}&=&
g^2am_{(13)}\frac{1}{2N}\left(T_S^{(m)}-T_P^{(m)}\right)
\nn\\&&
+g^2\left(-\left(C_F+\frac{1}{N}\right)T_A^{(p)}am_{(13)}
+\frac{1}{N}\left(T_S^{(p)}am_{(12)}-T_P^{(p)}am_{(13)}\right)\right),
\\
B'_{29}&=&
-g^2am_{(13)}\frac{1}{2}\left(T_S^{(m)}-T_P^{(m)}\right)
-g^2\left(T_S^{(p)}am_{(12)}-T_P^{(p)}am_{(13)}\right),
\\
B'_{2,10}&=&
g^2am_{(13)}\frac{1}{2N}\left(T_S^{(m)}-T_P^{(m)}\right)
+g^2\left(C_FT_A^{(p)}am_{(13)}
+\frac{1}{N}\left(T_S^{(p)}am_{(12)}-T_P^{(p)}am_{(13)}\right)\right),
\nn\\&&
\\
B_{31}&=&
-g^2am_{(12)}C_F\left(T_V^{(m)}-T_A^{(m)}\right)
+g^2\frac{1}{2N}T_A^{(p)}am_{(12)},
\\
B_{32}&=&
-g^2\frac{1}{2}T_A^{(p)}am_{(12)},
\\
B_{33}&=&
g^2aM\left(
C_F\left(\Sigma_1+\Sigma_1^{(1)}\right)
+\left(C_F+\frac{1}{2N}\right)\left(T_V^{(m)}+T_A^{(m)}\right)
-\frac{1}{2N}\left(T_S^{(m)}+T_P^{(m)}\right)\right)
\nn\\&&
-g^2\frac{1}{N}\left(T_S^{(p)}am_{(13)}+T_P^{(p)}aM\right),
\\
B_{34}&=&
-g^2aM\frac{1}{2}\left(T_V^{(m)}+T_A^{(m)}-T_S^{(m)}-T_P^{(m)}\right)
+g^2\left(T_S^{(p)}am_{(13)}+T_P^{(p)}aM\right),
\\
B'_{35}&=&
-g^2am_{(13)}\frac{1}{2N}\left(T_V^{(m)}-T_A^{(m)}\right)
-g^22C_FT_A^{(p)}am_{(13)},
\\
B'_{36}&=&
g^2am_{(13)}\frac{1}{2}\left(T_V^{(m)}-T_A^{(m)}\right),
\\
B'_{37}&=&
g^2am_{(14)}\frac{1}{2N}\left(T_S^{(m)}-T_P^{(m)}\right)
\nn\\&&
-g^2\left(
\left(2C_F-\frac{1}{2N}\right)T_A^{(p)}am_{(14)}
+\frac{1}{N}\left(T_P^{(p)}am_{(14)}-T_S^{(p)}am_{(12)}\right)\right),
\\
B'_{38}&=&
-g^2am_{(14)}\frac{1}{2}\left(T_S^{(m)}-T_P^{(m)}\right)
+g^2\left(-\frac{1}{2}T_A^{(p)}am_{(14)}+T_P^{(p)}am_{(14)}-T_S^{(p)}am_{(12)}
\right),
\\
B'_{39}&=&
-g^2am_{(14)}\frac{1}{2N}\left(T_S^{(m)}-T_P^{(m)}\right)
\nn\\&&
+g^2\left(\frac{1}{2N}T_A^{(p)}am_{(14)}
+\frac{1}{N}\left(T_P^{(p)}am_{(14)}-T_S^{(p)}am_{(12)}\right)\right),
\\
B'_{3,10}&=&
g^2am_{(14)}\frac{1}{2}\left(T_S^{(m)}-T_P^{(m)}\right)
-g^2\left(\frac{1}{2}T_A^{(p)}am_{(14)}
+\left(T_P^{(p)}am_{(14)}-T_S^{(p)}am_{(12)}\right)\right),
\\
B_{41}&=&
-g^2am_{(12)}\frac{1}{2}\left(T_V^{(m)}-T_A^{(m)}\right)
-g^2\frac{1}{2}T_A^{(p)}am_{(12)},
\\
B_{42}&=&
g^2am_{(12)}\frac{1}{2N}\left(T_V^{(m)}-T_A^{(m)}\right)
+g^2\frac{1}{2N}T_A^{(p)}am_{(12)},
\\
B_{43}&=&0,
\\
B_{44}&=&
g^2aMC_F\left(
\left(\Sigma_1+\Sigma_1^{(1)}\right)
+\left(T_S^{(m)}+T_P^{(m)}\right)\right)
+g^22C_F\left(T_S^{(p)}am_{(13)}+T_P^{(p)}aM\right),
\\
B'_{45}&=&
g^2am_{(13)}\frac{1}{2}\left(T_V^{(m)}-T_A^{(m)}\right)
-g^2T_A^{(p)}am_{(13)},
\\
B'_{46}&=&
-g^2am_{(13)}\frac{1}{2N}\left(T_V^{(m)}-T_A^{(m)}\right)
+g^2\frac{1}{N}T_A^{(p)}am_{(13)},
\\
B'_{47}&=&
-g^2\frac{3}{2}T_A^{(p)}am_{(14)},
\\
B'_{48}&=&
-g^2am_{(14)}C_F\left(T_S^{(m)}-T_P^{(m)}\right)
\nn\\&&
+g^2\left(\frac{3}{2N}T_A^{(p)}am_{(14)}
+2C_F\left(T_P^{(p)}am_{(14)}-T_S^{(p)}am_{(12)}\right)\right),
\\
B'_{49}&=&
-g^2\frac{1}{2}1T_A^{(p)}am_{(14)},
\\
B'_{4,10}&=&
g^2am_{(14)}C_F\left(T_S^{(m)}-T_P^{(m)}\right)
\nn\\&&
+g^2\left(\frac{1}{2N}T_A^{(p)}am_{(14)}
-2C_F\left(T_P^{(p)}am_{(14)}-T_S^{(p)}am_{(12)}\right)\right).
\end{eqnarray}
where the $g^2$ dependence is shown explicitly.
The quark masses used here is given by
\begin{eqnarray}
&&
M=\frac{1}{4}\left(m_{1}+m_2+m_{3}+m_4\right)_R,
\\&&
m_{(ij)}=\frac{1}{4}\left(m_{i}+m_j-m_{i'}-m_{j'}\right)_R,\quad
\{i',j'\}=\{1,2,3,4\}-\{i,j\}.
\end{eqnarray}

The $O(g^2ap)$ coefficients are given as
\begin{eqnarray}
&&
C_{11}=-g^2C_FT_V^{(p)},\quad
C_{12}=0,\quad
C_{13}=-g^2\frac{1}{2N}T_V^{(p)},\quad
C_{14}=g^2\frac{1}{2}T_V^{(p)},
\\&&
C_{21}=-g^2\frac{1}{2}T_V^{(p)},\quad
C_{22}=g^2\frac{1}{2N}T_V^{(p)},\quad
C_{23}=0,\quad
C_{24}=g^2C_FT_V^{(p)},
\\&&
C_{35}=g^2C_FT_V^{(p)},\quad
C_{36}=0,\quad
C_{37}=-g^2\frac{1}{2N}T_V^{(p)},\quad
C_{38}=g^2\frac{1}{2}T_V^{(p)},
\\&&
C_{45}=g^2\frac{1}{2}T_V^{(p)},\quad
C_{46}=-g^2\frac{1}{2N}T_V^{(p)},\quad
C_{47}=g^2\frac{1}{2}T_V^{(p)},\quad
C_{48}=-g^2\frac{1}{2N}T_V^{(p)}
\end{eqnarray}
with mixing operator vertex defined as
\begin{eqnarray}
\wt{\Gamma}_1&=&
1\wt{\otimes}1ia\left(p_1+p_2\right)_\nu
\biggl(
\left(\sigma_{\mu\nu} \otimes \gamma_\mu\gamma_5\right)
-\left(\gamma_\mu\gamma_5\otimes\sigma_{\mu\nu}\right)
\biggr),
\label{eqn:gamma-tilde-1}
\\
\wt{\Gamma}_2&=&
1\wt{\odot}1ia\left(p_1+p_2\right)_\nu
\biggl(
\left(\sigma_{\mu\nu} \otimes \gamma_\mu\gamma_5\right)
-\left(\gamma_\mu\gamma_5\otimes\sigma_{\mu\nu}\right)
\biggr),
\\
\wt{\Gamma}_3&=&
1\wt{\otimes}1ia\left(p_1+p_4\right)_\nu
\biggl(\left(\sigma_{\mu\nu}\odot\gamma_\mu\gamma_5\right)
-\left(\gamma_\mu\gamma_5\odot\sigma_{\mu\nu}\right)\biggr),
\\
\wt{\Gamma}_4&=&
1\wt{\odot}1ia\left(p_1+p_4\right)_\nu
\biggl(\left(\sigma_{\mu\nu}\odot\gamma_\mu\gamma_5\right)
-\left(\gamma_\mu\gamma_5\odot\sigma_{\mu\nu}\right)\biggr),
\\
\wt{\Gamma}_5&=&
1\wt{\otimes}1ia\left(p_1+p_2\right)_\nu\biggl(
\left(\sigma_{\mu\nu} \otimes \gamma_\mu\gamma_5\right)
+\left(\gamma_\mu\gamma_5\otimes\sigma_{\mu\nu}\right)
\biggr),
\\
\wt{\Gamma}_6&=&
1\wt{\odot}1ia\left(p_1+p_2\right)_\nu\biggl(
\left(\sigma_{\mu\nu} \otimes \gamma_\mu\gamma_5\right)
+\left(\gamma_\mu\gamma_5\otimes\sigma_{\mu\nu}\right)
\biggr),
\\
\wt{\Gamma}_7&=&
1\wt{\otimes}1ia\left(p_1+p_3\right)_\nu\biggl(
\left( \sigma_{\mu\nu}C^{-1}\circledast C\gamma_\mu\gamma_5\right)
-\left( \gamma_\mu\gamma_5C^{-1}\circledast C\sigma_{\mu\nu}\right)
\biggr),
\\
\wt{\Gamma}_8&=&
1\wt{\odot}1ia\left(p_1+p_3\right)_\nu\biggl(
\left( \sigma_{\mu\nu}C^{-1}\circledast C\gamma_\mu\gamma_5\right)
-\left( \gamma_\mu\gamma_5C^{-1}\circledast C\sigma_{\mu\nu}\right)
\biggr).
\label{eqn:gamma-tilde-8}
\end{eqnarray}

%\begin{references}

%\end{references}

\newpage
%%%%%%%%%%%%%%%%%%%%%%%%%%%%%%%%%%%%%%%%%%%%%%%%%%%%%%%%%%%%%%%%%%%%%
\begin{table}[ht]
\begin{center}
\caption{Finite part $V_\Gamma$ for bilinear operators \cite{Aoki:1998ar}. 
Coefficients of the term $c_{\rm SW}^n (n=0,1,2)$ are given in the column 
marked as $(n)$.
Terms proportional to $c_{\rm SW}^1$ are zero for pseudoscalar $P$.     
}
\label{tab:local}
\begin{tabular}{|ccc|ccc|ccc|ll|ccc|}
\hline
\multicolumn{3}{|c|}{$V$} &
\multicolumn{3}{c|}{$A$} &
\multicolumn{3}{c|}{$S$} &
\multicolumn{2}{c|}{$P$} & 
\multicolumn{3}{c|}{$T$} \\
(0) & (1) & (2)  &
(0) & (1) & (2)  &
(0) & (1) & (2)  &
(0) & (2)  &
(0) & (1) & (2)  \\
\hline
$6.275$&$-1.725$&$ 0.637$& $3.367$& $1.725$&$-0.637$&
$2.533$&$ 6.902$&$-0.293$& $8.348$& $2.254$&$ 4.615$&$-1.150$&$-0.327$ \\
\hline
\end{tabular}
\end{center}
\end{table}

\begin{table}[ht]
\caption{Finite constants for quark self-energy \cite{Aoki:1998ar}. 
Coefficients of the term $c_{\rm SW}^n (n=0,1,2)$ are given in the column 
marked as $(n)$.
Tadpole contribution is also listed.}
\label{tab:self}
\begin{center}
\begin{tabular}{|cccc|}
\hline
\multicolumn{4}{|c|}{$\Sigma_1$} \\
(0) & tad& (1) & (2)  \\
\hline
$4.825$ & $7.482$ & $-1.601$ & $-0.973$ \\
\hline
\end{tabular}
\end{center}
\end{table}

\begin{table}[ht]
\begin{center}
\caption{Finite part $z^g_{ij}$ of the renormalization factor from gluon
exchanging diagrams.
The DRED scheme is adopted.
The color factor is set to $N=3$.
Coefficients of the term $c_{\rm SW}^n (n=0,1,2)$ are given in the column 
marked as $(n)$.
}
\label{tab:gluon-exchanging}
\begin{tabular}{|ccc|ccc|ccc|ccc|}
\hline
\multicolumn{3}{|c|}{$z^g_{11}$} &
\multicolumn{3}{c|}{$z^g_{55}$} &
\multicolumn{3}{c|}{$z^g_{66}$} &
\multicolumn{3}{c|}{$z^g_{12}$} \\
(0) & (1) & (2)  &
(0) & (1) & (2)  &
(0) & (1) & (2)  &
(0) & (1) & (2)  \\
\hline
$-23.596$ & $3.119$ & $2.268$ &
$-25.183$ & $5.420$ & $2.922$ &
$-18.041$ & $-4.933$ & $-0.020$ &
$ -2.381$ & $0.451$ & $-2.020$ \\
\hline
\end{tabular}
\end{center}
\end{table}

\begin{table}[ht]
\begin{center}
\caption{Finite part $z^g_{ij}$ of the renormalization factor from gluon
exchanging diagrams in the NDR scheme.
The color factor is set to $N=3$.
$c_{\rm SW}$ dependent terms are the same as that in the DRED scheme
$(n=1,2)$.
}
\label{tab:gluon-exchanging-NDR}
\begin{tabular}{|c|c|c|c|c|c|}
\hline
{$z^{g(0)}_{11}$} & {$z^{g(0)}_{55}$} & {$z^{g(0)}_{66}$} &
{$z^{g(0)}_{12}$} & {$z^{g(0)}_{56}$} & {$z^{g(0)}_{65}$} \\
\hline
$-24.096$ & $-25.350$ & $-19.708$ &
$ -4.881$ & $ -1.120$ & $-3$ \\
\hline
\end{tabular}
\end{center}
\end{table}

\begin{table}[ht]
\caption{Finite part in the penguin diagram contribution on the lattice.
Coefficients of the term $c_{\rm SW}^n (n=0,1)$ are given in the column 
marked as $(n)$.}
\label{tab:penguin-lat}
\begin{center}
\begin{tabular}{|cc|}
\hline
\multicolumn{2}{|c|}{$V_{\rm pen}$} \\
(0) & (1) \\
\hline
$-1.7128$ & $-1.0878$ \\
\hline
\end{tabular}
\end{center}
\end{table}

\begin{table}[ht]
\caption{Finite part in the renormalization factor for the penguin
diagram contribution.
Coefficients of the term $c_{\rm SW}^k (k=0,1)$ are given in the column 
marked as $(k)$.}
\label{tab:penguin}
\begin{center}
\begin{tabular}{|c|ccc|c|}
\hline
$z^{\rm pen}_i({\rm DRED})^{(0)}$ &
$z^{\rm pen}_2({\rm NDR})^{(0)}$ &
$z^{\rm pen}_{2n-1}({\rm NDR})^{(0)}$ &
$z^{\rm pen}_{2n}({\rm NDR})^{(0)}$ &
$(z^{\rm pen}_i)^{(1)}$ \\
\hline
$-0.2039$ &
$1.0462$ &
$1.0462$ &
$0.0461$ &
$1.0878$ \\
\hline
\end{tabular}
\end{center}
\end{table}

\begin{table}[ht]
\begin{center}
\caption{Finite part $z^{g({\rm MF})}_{ii}$ of the renormalization
factor from gluon exchanging diagrams for the mean field improvement
given with the tadpole subtraction in the DRED and NDR scheme.
}
\label{tab:mena-field}
\begin{tabular}{|c|c|c|c|}
\hline
scheme & {$z^{g{\rm (MF)}(0)}_{11}$} & {$z^{g{\rm (MF)}(0)}_{55}$}
& {$z^{g{\rm (MF)}(0)}_{66}$} \\
\hline
DRED & $-3.644$ & $-5.231$ & $1.911$ \\
NDR  & $-4.144$ & $-5.398$ & $0.244$ \\
\hline
\end{tabular}
\end{center}
\end{table}

\begin{table}[ht]
\caption{Coefficients of the $O(g^2ap)$ correction to bilinear operators
 for the Iwasaki gauge action.
 The improvement coefficient is set to its tree level value
 $c_{\rm SW}=1$.}
\label{tab:bilinear-oap}
\begin{center}
\begin{tabular}{|cccc|}
\hline
%gauge action $c_1$ &
$(16\pi^2)T_A^{(p)}$ &
$(16\pi^2)T_V^{(p)}$ &
$(16\pi^2)T_P^{(p)}$ &
$(16\pi^2)T_S^{(p)}$ \\
\hline
%$-0.331$ &
$0.4519(22)$ &
$-1.1478(67)$ &
$0$ &
$0$ \\
\hline
\end{tabular}
\end{center}
\end{table}

\begin{table}[ht]
\caption{Coefficients of the $O(g^2am)$ correction to bilinear operators
 for the Iwasaki gauge action.
 The improvement coefficient is set to its tree level value
 $c_{\rm SW}=1$.}
\label{tab:bilinear-oam}
\begin{center}
\begin{tabular}{|cccc|}
\hline
%gauge action $c_1$ &
$(16\pi^2)T_A^{(m)}$ &
$(16\pi^2)T_V^{(m)}$ &
$(16\pi^2)T_P^{(m)}$ &
$(16\pi^2)T_S^{(m)}$ \\
\hline
%$-0.331$ &
$-0.9764(47)$ &
$-0.995(13)$ &
$-1.17592(48)$ &
$-4.1201(36)$ \\
\hline
\end{tabular}
\end{center}
\end{table}

\begin{table}[ht]
\caption{Coefficients of the $O(g^2)$ and $O(g^2a)$ correction to the
 quark propagator for the Iwasaki gauge action.
 The improvement coefficient is set to its tree level value
 $c_{\rm SW}=1$.}
\label{tab:quark-oa}
\begin{center}
\begin{tabular}{|c|ccc|}
\hline
gauge action $c_1$ &
$(16\pi^2)\Sigma_1$ &
$(16\pi^2)\Sigma_1^{(1)}$ &
$(16\pi^2)z_m$ \\
\hline
$-0.331$ &
$2.25022(44)$ &
$-9.9266(69)$ &
$-11.395(53)$ \\
\hline
\end{tabular}
\end{center}
\end{table}

%%%%%%%%%%%%%%%%%%%%% FIGURES %%%%%%%%%%%%%%%%%%%%%%%%%%%

\end{document}